\documentclass[twocolumn, twocolappendix]{aastex631}

\shorttitle{Supernova Implosion}
\shortauthors{Romano et al.}

\begin{document}

\title{Cloud Formation by Supernova Implosion}

\correspondingauthor{Leonard E. C. Romano}
\email{lromano@usm.lmu.de}

\author[0000-0001-8404-3507]{Leonard E. C. Romano}
\affiliation{Universitäts-Sternwarte, Fakultät für Physik, Ludwig-Maximilians-Universität München, Scheinerstr. 1, D-81679 München, Germany}
\affiliation{Max-Planck-Institut für extraterrestrische Physik, Giessenbachstr. 1, D-85741 Garching, Germany}
\affiliation{Excellence Cluster ORIGINS, Boltzmannstr. 2, D-85748 Garching, Germany}

\author[0000-0002-8759-941X]{Manuel Behrendt}
\affiliation{Universitäts-Sternwarte, Fakultät für Physik, Ludwig-Maximilians-Universität München, Scheinerstr. 1, D-81679 München, Germany}
\affiliation{Max-Planck-Institut für extraterrestrische Physik, Giessenbachstr. 1, D-85741 Garching, Germany}

\author{Andreas Burkert}
\affiliation{Universitäts-Sternwarte, Fakultät für Physik, Ludwig-Maximilians-Universität München, Scheinerstr. 1, D-81679 München, Germany}
\affiliation{Max-Planck-Institut für extraterrestrische Physik, Giessenbachstr. 1, D-85741 Garching, Germany}
\affiliation{Excellence Cluster ORIGINS, Boltzmannstr. 2, D-85748 Garching, Germany}



\begin{abstract}
The deposition of energy and momentum by supernova explosions has been subject to numerous studies in the past few decades. However, while there has been some work that focused on the transition from the adiabatic to the radiative stage of a supernova remnant (SNR), the late radiative stage and merging with the interstellar medium (ISM) have received little attention. Here, we use three-dimensional, hydrodynamic simulations, focusing on the evolution of SNRs during the radiative phase, considering a wide range of physical explosion parameters ($n_{\text{H, ISM}} \in \left[0.1, 100\right] \text{cm}^{-3}$ and $E_{\text{SN}} \in \left[1, 14\right]\times 10^{51}\, \text{erg}$).
We find that the radiative phase can be subdivided in four stages: A pressure driven snowplow phase during which the hot overpressurized bubble gas is evacuated and pushed into the cold shell, a momentum conserving snowplow phase which is accompanied by a broadening of the shell, an implosion phase where cold material from the back of the shell is flooding the central vacuum and a final cloud phase, during which the imploding gas is settling as a central, compact overdensity. 
The launching timescale for the implosion ranges from a few 100 kyr to a few Myr, while the cloud formation timescale ranges from a few to about 10 Myr.
The highly chemically enriched clouds can become massive ($M_{\text{cl}} \sim 10^3 \, - \, 10^4 \, \text{M}_{\odot}$) and self-gravitating within a few Myr after their formation, providing an attractive, novel pathway for supernova induced star and planet formation in the ISM.
\end{abstract}



\section{Introduction} \label{sec:intro}

It has long been recognized that supernovae (SNe) play an important role in maintaining the balance and structure of the interstellar medium (ISM). Even though there is only about one supernova per 100 M$_{\odot}$ of formed stars, due to their enormous energy output, these destructive events can have an enormous impact on their surroundings. 

Many aspects of galaxy formation and evolution, like star formation and the modeling of galactic outflows \citep[e.g.][]{2017MNRAS.470L..39F, 2022ApJ...932...88O}, are tightly linked to the evolution of supernova remnants (SNRs). 
SNe are believed to maintain the hot phase \citep[e.g.][]{2004A&A...425..899D,2023MNRAS.523.6336B}, drive turbulence and outflows \citep[e.g.][]{1995ApJ...440..634R, 2018MNRAS.477.2716K, 2018MNRAS.481.3325F, 2022ApJS..262....9O}, regulate the star formation rate in disk galaxies \citep[e.g.][]{2012ApJ...754....2S, 2019MNRAS.484.2632S, 2023MNRAS.521.5712H} and enrich their surroundings with heavy elements and dust \citep[e.g.][]{1989ApJ...344..325K, 2007MNRAS.378..973B, 2007ApJ...666..955N}.

Of particular interest is the concept of positive SN feedback or \textit{triggered star formation}, 
where a strong shock wave compresses the gas, leading to further collapse, fragmentation and eventually the formation of new stars. This processes has been predicted in numerous theoretical works \citep[e.g.][]{2017ApJ...851..147D, 2018A&A...619A.120K, 2023MNRAS.521.5712H} and has recently been confirmed with observations from the \textit{Gaia} mission \citep[e.g.][]{2022Natur.601..334Z, 2022A&A...667A.163M, 2023arXiv230207853R}. 

SNR evolution in a uniform medium has been studied at great length using analytical models \citep[e.g.][]{1972ARA&A..10..129W, 1978ApJ...225..442G, 1988RvMP...60....1O}, and numerical simulations in one \citep[e.g.][]{1974ApJ...188..501C, 1988ApJ...334..252C, 2016MNRAS.456..710F}, two \citep[e.g.][]{1998ApJ...500..342B, 2011ApJ...731...13N, 2023MNRAS.521.5354M} and three dimensions \citep[e.g.][]{2015ApJ...802...99K, 2023MNRAS.523.1421M}, which have lead to a comprehensive picture comprised of a series of different stages, characterized by different deceleration parameters $q = - \text{d}^2R / \text{d}t^2$ and conserved quantities. 
In the first stage, known as the \textit{free expansion} phase the SNR expands with constant velocity until the reverse shock has fully thermalized the ejecta \citep{1999ApJS..120..299T}. 
In the next so-called Sedov-Taylor (ST) phase \citep{1959sdmm.book.....S, 1950RSPSA.201..159T}, the SNR expands adiabatically as cooling losses are still negligible and thus energy is conserved. 
The ST phase ends, when radiative cooling losses become important and a thin, cold shell forms at the shock front. 
After the shell has formed, the SNR keeps expanding in what is known as the pressure-driven snowplow (PDS) phase \citep{1982ApJ...253..268C, 1988RvMP...60....1O}. 
During the PDS, the hot bubble is rapidly evacuated as hot material is pushed into the shell \citep{1983ApJ...273..267G, 1988ApJ...334..252C, 2015ApJ...802...99K}. 
Once the bubble pressure has dropped below that of the shell, the PDS ends and transitions into a momentum conserving snowplow (MCS) phase \citep{1988ApJ...334..252C, 1998ApJ...500...95T}. 
It has been claimed that the SNR evolution ends, when the shock velocity becomes comparable to the typical velocity dispersion of the ambient medium \citep{1988ApJ...334..252C, 2011piim.book.....D, 2013MNRAS.433.1970F, 2018MNRAS.477.2716K} and the shock merges with the ISM. 
However, the details of the merging have only received little attention and it remains unclear how the evacuated bubble carved out by the blast wave is refilled.

In order to address this gap, in this work, we utilize three-dimensional, hydrodynamical simulations with cooling to study the late radiative stage of SNR evolution and the onset of fade out. 
We thus provide a more complete picture for the later stages of SNR evolution. 
We show that as the SNR shell pressure approaches the ISM pressure, the SNR implodes, filling the central cavity. 
This implosion leads to the formation of a dense compact, cloud in its center, which has the potential to form new stars and thus provides a novel pathway for triggered star formation. 

The remainder of this paper is organized as follows. In section \ref{sec:methods} we describe the numerical scheme and the setup of our simulation suite. In section \ref{sec:results} we present the results of our numerical simulations. We discuss the limitations and implications of our results in section \ref{sec:discussion}. Finally, we summarize our findings and conclude in section \ref{sec:summary}. In the appendix we present a number of tests, related to the question of numerical convergence and the adopted treatment for radiative cooling.

\begin{deluxetable*}{lccccc}
\tablecaption{Overview of the simulation suite}\label{tab:models}
\tablewidth{0.9\textwidth}
\tablehead{\colhead{Model} & \colhead{$N_{\text{SN}}$} & \colhead{$n_{\text{H}}$} & \colhead{$\Delta x_{\text{min}}$} & \colhead{$\lfloor R_{\text{sf}} \, / \, \Delta x_{\text{min}} \rfloor^1$} & \colhead{Comment} \\ 
\colhead{} & \colhead{} & \colhead{$\left[\text{cm}^{-3}\right]$} & \colhead{$\left[\text{pc}\right]$}  & \colhead{} & \colhead{}} 
\startdata
N1\_n-1\_L11    & 1  & 0.1 & 0.5    & 118&\\
N1\_n0\_L11     & 1  & 1   & 0.5    & 45&\\
N1\_n0\_L11\_HC & 1  & 1   & 0.5    & 45& L = 256 pc, $\Delta x_{\text{max}} = 2 \, \text{pc}$, $t_{\text{end}} = 1.5 \, \text{Myr}$\\
N1\_n1\_L11     & 1  & 10  & 0.5    & 17&\\
N1\_n1\_L11\_HC & 1  & 10  & 0.5    & 17& L = 64 pc, $\Delta x_{\text{max}} = 0.5 \, \text{pc}$, $t_{\text{end}} = 1.5 \, \text{Myr}$\\
N1\_n2\_L12     & 1  & 100 & 0.25   & 13&\\
N1\_n2\_L13     & 1  & 100 & 0.125  & 26&\\
N1\_n2\_L13\_noAMR & 1  & 100   & 0.125    & 26& L = 128 pc, $\Delta x_{\text{max}} = 0.125 \, \text{pc}$\\
N1\_n2\_L13\_HC & 1  & 100 & 0.125  & 26& L = 64 pc, $\Delta x_{\text{max}} = 0.5 \, \text{pc}$, $t_{\text{end}} = 1.5 \, \text{Myr}$\\
N1\_n2\_L14     & 1  & 100 & 0.0625 & 52&\\
N5\_n-1\_L11    & 5  & 0.1 & 0.5    & 189&\\
N5\_n0\_L11     & 5  & 1   & 0.5    & 72&\\
N5\_n1\_L11     & 5  & 10  & 0.5    & 27&\\
N5\_n2\_L11     & 5  & 100 & 0.5    & 10&\\
N14\_n-1\_L11   & 14 & 0.1 & 0.5    & 255&\\
N14\_n0\_L11    & 14 & 1   & 0.5    & 97&\\
N14\_n1\_L11    & 14 & 10  & 0.5    & 36&\\
N14\_n2\_L11    & 14 & 100 & 0.5    & 14&\\
N1\_n1\_L11\_Dust     & 1  & 10  & 0.5    & 17& Dust only cooling model of \citet{2020MNRAS.497.4857P}.\\
N1\_n1\_L11\_PS20     & 1  & 10  & 0.5    & 17& Fiducial cooling model of \citet{2020MNRAS.497.4857P}.\\
\enddata
\tablenotetext{1}{Equation \ref{eq:r_sf}}
\end{deluxetable*}

\section{Methods} \label{sec:methods}
\subsection{Numerical Methods} \label{sec:numerics}

We utilize the adaptive mesh refinement (AMR) code {\sc ramses} \citep{2002A&A...385..337T} to simulate the hydrodynamic evolution of blast waves in a uniform density medium, including radiative cooling. 
{\sc ramses} is solving the system of hydrodynamic equations utilizing a second-order unsplit Godunov method (MUSCL scheme) on a finite volume, cartesian grid. 
Variables at the cell interfaces are reconstructed from the cell-centered values using the HLLC Riemann solver \citep{1994ShWav...4...25T} with MinMod total variation diminishing scheme. Cooling is solved for the default \textit{courty} cooling function implemented in {\sc ramses}, which provides a basic treatment of primordial chemistry, metal line cooling and heating due to ultraviolet background (UVB) radiation.

In our fiducial simulation suite we consider a cubic computational domain with a side length of $L = 1024\, \text{pc}$ and periodic boundaries, which we refine with $l_{\text{min}} = 7$ to $l_{\text{max}} = 11$ refinement levels, corresponding to a spatial resolution of $\Delta x_{\text{max}} = 8 \, \text{pc}$ and $\Delta x_{\text{min}} = 0.5 \, \text{pc}$, respectively. However, we ensure that the resolution criterion of \citet{2015ApJ...802...99K} is fulfilled and accordingly increase the resolution in runs, where the expected radius at shell formation would not be resolved with at least 10 grid cells. 

Initially, the simulation domain is filled with uniform density gas with $\log n_{\text{H}} \, \left[\text{cm}^{-3}\right] \in \{-1, 0, 1, 2\}$ at solar metallicity and an initial temperature set to be close to cooling equilibrium. 
In the domain center, we initialize the explosive ejecta uniformly within a spherical region of radius $R_{\text{inj}} \sim 5 \, \Delta x_{\text{min}}$. 
We inject $E_{\text{SN}} = 10^{51} \, \text{erg}$ and $M_{\text{ej}} = 5 \, \text{M}_{\odot}$ per SN, corresponding to an initial ejecta temperature of $T_{\text{SN}} \sim 10^9 \, K$. 
We ensure that the injection region is maximally refined, by statically refining the central $R_{\text{ref}} = 50 \, \Delta x_{\text{min}}$ with the maximum resolution.

In order to ensure that the shock and the bubble are maximally refined, while only as little as possible of the surrounding medium is refined, we advect a passive scalar variable $Z_{\text{ej}}$ with the injected mass and maximally refine all cells where $Z_{\text{ej}} > 10^{-15}$. 
The criterion might fail, if numerical errors in the advection of pristine cells trigger the criterion or if cells just behind the shock are not polluted enough to trigger refinement. 
However, we have checked that both of these cases do not occur frequently enough to cause any serious problems. 
In the Appendix \ref{app:amr} we discuss the role of the AMR in more detail.

Besides different densities we also consider different explosion strengths, mimicking the feedback from a single stellar population with $N_{\text{SN}} \in \{1, 5, 14\}$ massive stars exploding all at once, by simply injecting $N_{\text{SN}}$ times as much mass and energy. We thus label a model with $N_{\text{SN}} = x$, $\log n_{\text{H}} = y$ and $l_{\text{max}} = z$ as $\text{N}x\_\text{n}y\_\text{L}z$. While this simplistic approach can capture some aspects of clustered feedback, it is worth noting that studies that take into account the time delay between explosions find some qualitative differences, such as an increased momentum per SN \citep{2015MNRAS.451.2757W, 2019MNRAS.483.3647G} and a longer lived hot bubble \citep{2017ApJ...834...25K}.

All models are run until $t = 14 \, \text{Myr}$ at which point the largest bubbles are reaching the domain limits. 
We reran some of the models in a smaller domain for a shorter time span, but with a much higher frequency of snapshots. 
For these models we add the suffix \_HC to the name and the part of the name referring to the resolution refers to the \textit{equivalent} refinement level for the fiducial domain, i.e. L11 refers to $\Delta x_{\text{min}} = 0.5 \, \text{pc}$ in both the fiducial and the HC runs.

In appendix \ref{app:cooling} we will discuss the effect that different cooling functions could have on our results. 
To this end, we rerun a few of the models with different cooling tables taken from \citet{2020MNRAS.497.4857P} and label the models with suffices corresponding to the respective alternate cooling model.

Finally, in section \ref{app:convergence} we are discussing the results obtained for the N1\_n2 model at different resolutions.

A list of all the different models and their properties is given in table \ref{tab:models}.

\subsection{Data Analysis} \label{sec:analysis}

\begin{figure}
\centering
\includegraphics[width=0.45\textwidth]{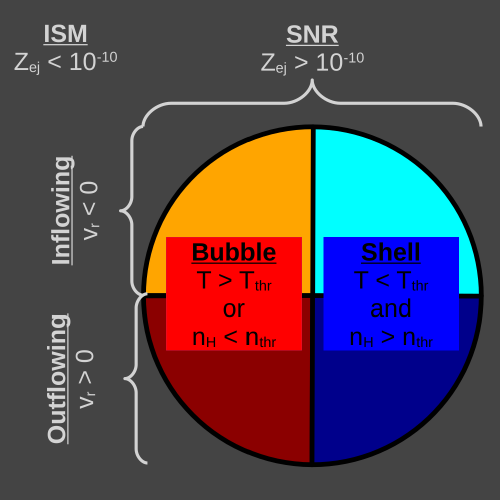}
\caption{Schematic overview of the different gas components, described in section \ref{sec:analysis}. The classification differentiates between the ISM (gray) and the SNR (color), consisting of a bubble (shades of red) and a shell (shades of blue), which itself might be out- or inflowing.}\label{fig:classification}
\end{figure}

In order to quantify the global evolution of the gas, we distinguish between bubble and shell gas and further within these components, differentiate between gas which is moving radially outward ($v_{r} > 0$) and inward ($v_{r} < 0$). 
We distinguish between the SNR and the ISM using the passive scalar, i.e. gas with $Z_{\text{ej}} > 10^{-10}$ is considered part of the SNR. 
The bubble is defined as SNR gas that is either hot ($T > 2 \times 10^4 \, \text{K}$) or diffuse ($n_{\text{H}} < 10^{-2} \, \text{cm}^{-3}$), while the shell is all SNR gas that is not part of the bubble. A summary of the classification is given in Figure \ref{fig:classification}.

\citet{2015ApJ...802...99K} use a similar criterion for the bubble gas, considering only the temperature of the gas. 
The addition of the density criterion only becomes important at late times, when the bubble has cooled below $10^4 \, K$ at which point the temperature criterion alone would fail to differentiate between the bubble and the shell.

The partition of the SNR into a bubble and shell, allows us to measure the shell formation timescale $t_{\text{sf}}$, which denotes the time when the cold shell at the shock front forms. 
We follow \citet{2015ApJ...802...99K}, who define the numerically measured shell formation timescale $t_{\text{sf}}^{\text{n}}$ as the time at which the mass of the hot bubble reaches its maximum.

\section{Results}\label{sec:results}

\begin{figure*}
\centering
\includegraphics[width=0.8\textwidth]{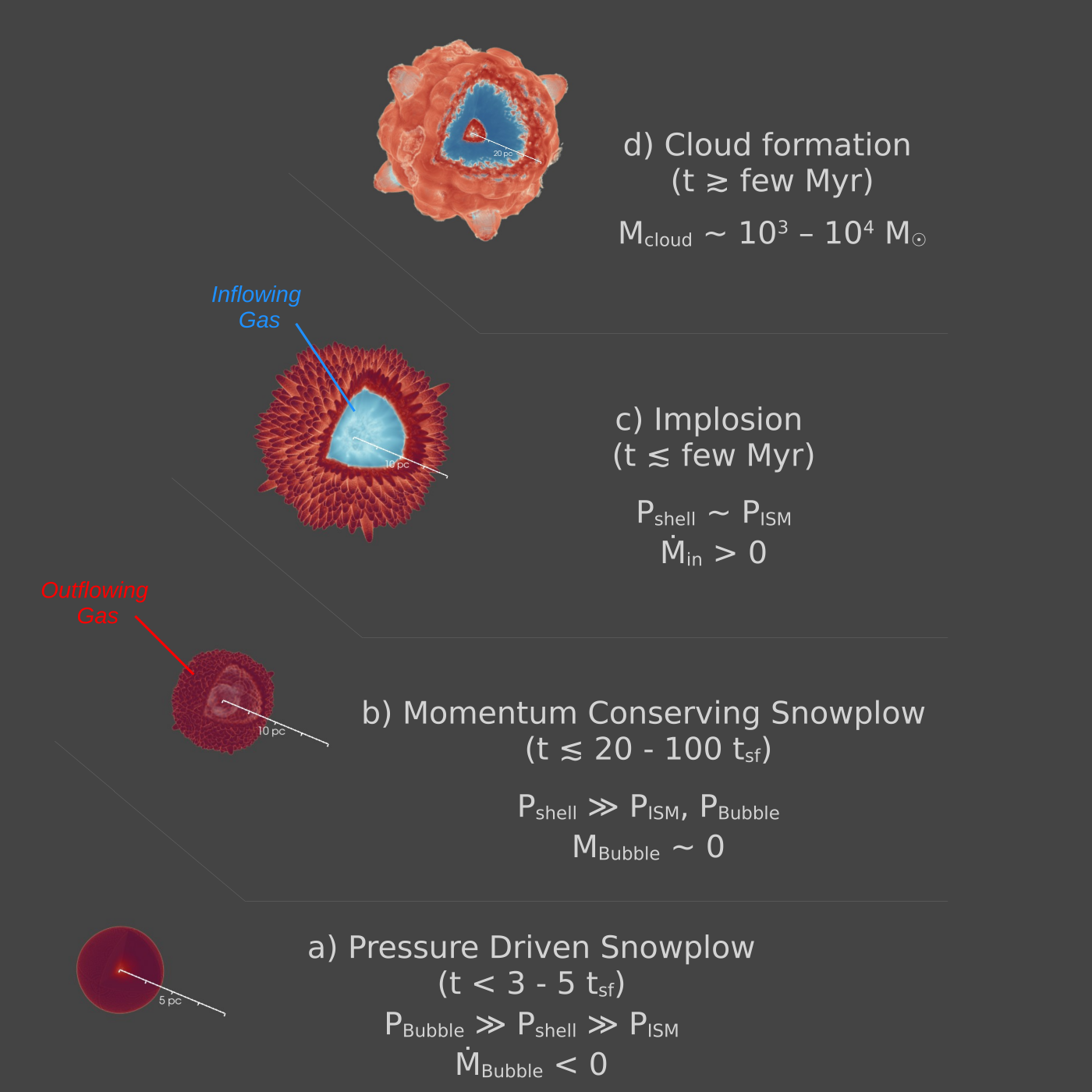}
\caption{Schematic overview of our proposed cloud formation mechanism. We show volume renderings of the radial mass flux in the model N1\_n2\_L14, during different phases of SNR late stage evolution. The physical scale differs between the frames. Opacity alpha is scaled with the logarithm of the density and is set to zero for densities in the range $n_{\text{H}} \in \left(95, 105\right) \text{cm}^{-3}$ in order to remove the foreground. The octant facing the observer has been made transparent.}\label{fig:money_plot}
\end{figure*}

In this section we describe a mechanism for the formation of a cloud through the implosion of a radiative SNR.
In section \ref{sec:overview} we give a brief overview of the physical mechanism. 
We provide a detailed description of our simulation results for a single SN in a high density ISM in section \ref{sec:N1_n2}.
In section \ref{sec:universality} we extend our analysis to the whole simulation suite and investigate the dependence of the relevant timescales on the explosion parameters in section \ref{sec:timescales}.
We describe the properties of the implosion clouds in section \ref{sec:cloud_properties}.
Finally, in section \ref{sec:model} we describe a model for the launching of the implosion.

\newpage
\subsection{Schematic Overview}\label{sec:overview}

Figure \ref{fig:money_plot} gives a schematic description of the evolution of the SNR after shell formation, which is separated into four stages.

In the first stage, shortly after shell formation the interior of the SNR is still hot and overpressurized relative to the isothermal shell, which is kept at about $T\sim 10^{4} \, \text{K}$. 
During this stage the rarefied bubble material tries to expand and as a result is pressed into the shell. 
This phase corresponds to the modified PDS phase described by \citet{1988ApJ...334..252C} and \citet{2015ApJ...802...99K}. 
After about $3\, - \,5$ shell formation timescales the bubble has been evacuated and its pressure has dropped below that of the shell.

At this point, the expansion of the SNR is entirely inertia driven, corresponding to the MCS.
The mass of the evacuated bubble is negligibly small and its density is many orders of magnitude below the ambient density. 
The shell which is now overpressurized relative to both the ISM and the bubble, begins to broaden, leading to a gradual reduction in the shell pressure and an accelerated weakening of the shock.
Meanwhile, the shell begins to get deformed and fragmented due to thin-shell overstability and nonlinear thin-shell instability \citep{1983ApJ...274..152V, 1998ApJ...500..342B}. 

Once the shell pressure reaches pressure equilibrium with the ISM a reflected version of the outgoing shock wave is launched driving cooling material from the shell back into the center, refilling the warm, evacuated cavity. 
We refer to this reflected wave as \textit{Implosion} or \textit{Backflow}. 
The implosion is very similar to the so-called "negative phase" in the context of terrestrial blast waves \citep[see e.g.][]{1977enw..book.....G} and appears to be a purely hydrodynamic realization of the hydromagnetic Rayleigh-Taylor instability (RTI) described by \citet{2000A&A...361..303B}.

After a few Myr, the backflow reaches the center of the SNR and collides with the backflowing gas from all directions. The colliding gas piles up in the center and is reflected, forming a slowly expanding cloud. The cloud keeps accreting material from the backflowing gas reaching a mass of $10^3 \, - \, 10^4 \, \text{M}_{\odot}$ within $\sim 10\, \text{Myr}$. 
As the cloud is directly formed from the SN ejecta, it is highly chemically enriched.

\subsection{Model for the Launching of the Backflow}\label{sec:model}

In the previous subsection we have given an overview of the different phases of the radiative stage. 
Here we describe a model for estimating the relevant timescales.

Right after shell formation, we assume that the bubble is following the modified PDS described by \citet{2015ApJ...802...99K}. In their description the thermal energy of the bubble evolves as
\begin{equation}
    E_{\text{th, PDS}} = 0.8 \, E_{\text{th, ST}} \left(\frac{R_{\text{sf}}}{R}\right)^2 \frac{t_{\text{sf}}}{t},
\end{equation}
where $t_{\text{sf}}$ and $R_{\text{sf}}$ are given by \citep{2015ApJ...802...99K}
\begin{eqnarray}
     t_{\text{sf}} \sim 0.044 \, E_{51}^{0.22} \, n_{0}^{-0.55} \,\text{Myr}, \label{eq:t_sf}\\
     R_{\text{sf}} \sim 22.6  \, E_{51}^{0.29} \, n_{0}^{-0.42} \,\text{pc},  \label{eq:r_sf}
\end{eqnarray}
$E_{\text{th, ST}} = 0.72 \, E_{\text{SN}}$ and 
\begin{equation}
    R = R_{\text{sf}} \left(\frac{t}{t_{\text{sf}}}\right)^{2/7}.
\end{equation}
Here $E_{51} = E_{\text{SN}} / \left(10^{51} \, \text{erg}\right)$.
The average bubble pressure is then given by
\begin{equation}
    P_{\text{Bubble, PDS}} = \left(\gamma - 1\right) \frac{E_{\text{th, PDS}}}{4\pi/3 R^3}.
\end{equation}
Meanwhile, the temperature of the shell remains roughly constant at $T_{\text{shell}} \sim 10^4 \, K$ and the compression ratio of the shell is $\chi \sim 10$, leading to a shell pressure of $P_{\text{shell, PDS}} \sim 10^5 \, n_{0} \, k_{\text{B}} \, \text{K} \, \text{cm}^{-3}$.
The PDS phase ends, when the pressure in the shell and bubble is equal, at
\begin{equation}\label{eq:t_PDS}
    t_{\text{PDS}} \sim 3.4 \, E_{51}^{0.05} \, n_{0}^{0.11} t_{\text{sf}} \sim 0.15 \, E_{51}^{0.27} \, n_{0}^{-0.44} \, \text{Myr}.
\end{equation}
The radius of the SNR at this time is
\begin{equation}\label{eq:R_PDS}
    R_{\text{PDS}} \sim 32.1 \, E_{51}^{0.3} \, n_{0}^{-0.39} \, \text{pc}.
\end{equation}

After the PDS phase has ended, the momentum of the shell remains constant and the MCS phase begins. 
During this phase the radius of the SNR evolves $\propto t^{1/4}$ and correspondingly the shock velocity evolves $\propto t^{-3/4}$. 
If one assumes a constant compression ratio and that the temperature of the shell is proportional to the square of the velocity, as one would expect for a strong shock, one finds for the pressure during the MCS phase
\begin{equation}
    P_{\text{Shell, MCS}} = P_{\text{shell, PDS}} \left(\frac{t}{t_{\text{PDS}}}\right)^{-3/2}.
\end{equation}
The SNR implodes, when the pressure of the shell approaches the pressure of the ISM. 
In the standard RAMSES cooling prescription at solar metallicity, which assumes collisional ionization equilibrium the pressure on the cooling-equilibrium curve for a given density is
approximately (see e.g. Figure 17 of \citet{2023ApJS..264...10K})
\begin{equation}
    P_{\text{ISM, eq}} \sim 6 \times 10^3 \, n_{0}^{1/2} \,  k_{\text{B}} \, \text{K} \, \text{cm}^{-3}.
\end{equation}
Thus, the launching timescale can be inferred as
\begin{equation}\label{eq:t_launch}
    t_{\text{launch}} \sim 0.98 \, E_{51}^{0.27} \, n_{0}^{-0.11} \, \text{Myr},
\end{equation}
and the radius of the SNR at this time is
\begin{equation}\label{eq:R_launch}
    R_{\text{launch}} \sim 51.3 \, E_{51}^{0.3} \, n_{0}^{-0.27} \, \text{pc}.
\end{equation}

Finally, cloud formation happens once the imploding shell has reached the center. 
The cloud formation timescale is thus the combination of the launching and the crossing timescale
\begin{equation}\label{eq:t_cf}
    t_{\text{cf}} \sim t_{\text{launch}} + \frac{R_{\text{launch}}}{V_{\text{in}}},
\end{equation}
where $V_{\text{in}}$ is a characteristic inflow velocity. The results in section \ref{sec:timescales} indicate that this velocity is independent of the explosion energy, but depends on the ambient density in a complicated way.

\subsection{Single Supernova in a High Density Medium}\label{sec:N1_n2}

\begin{figure}
\centering
\includegraphics[width=0.4\textwidth]{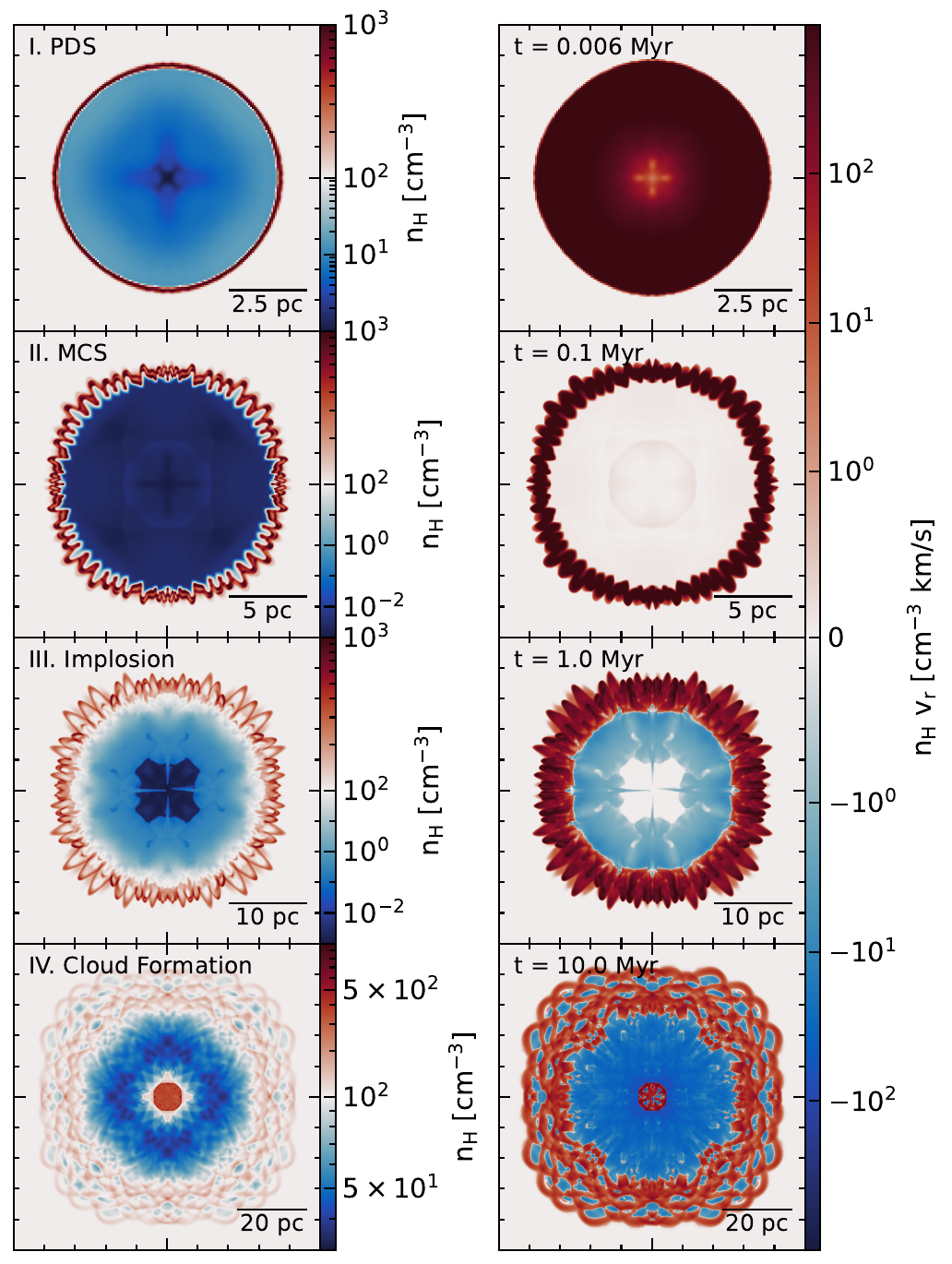}
\caption{Slices through the XY-plane of the model N1\_n2\_L14. The left and right columns display density and radial mass flux at different stages during the radiative phase of SNR evolution, respectively. The panels from top to bottom correspond to t = 0.006, 0.1, 1 and 10 Myr. Each panel showing density has a different color scale due to the large changes in dynamic range. The color scale for the density slices is logarithmic and asymmetrically centered around the ambient density $n_{\text{H, ISM}} = 100 \, \text{cm}^{-3}$. The color scale for the radial mass flux is the same in all panels.}\label{fig:slice}
\end{figure}

\begin{figure}
\centering
\includegraphics[width=0.45\textwidth]{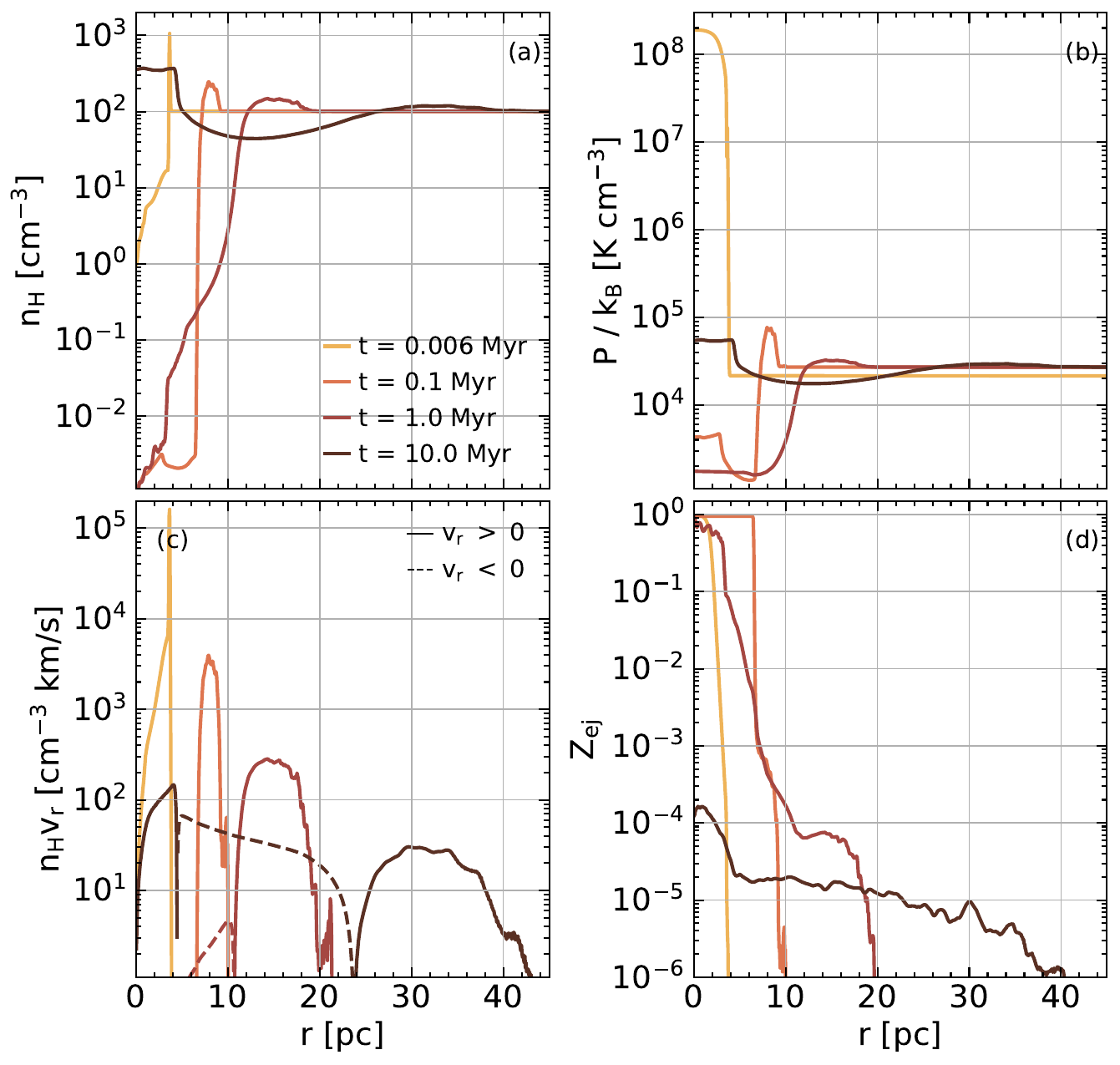}
\caption{Radial profiles of (a) number density, (b) pressure, (c) mass flux, and (d) enrichment of model N1\_n2\_L14. The differently shaded curves correspond to t = 0.006, 0.1, 1 and 10 Myr, respectively. In panel (c) solid (dashed) lines correspond to outward (inward) mass flux. }\label{fig:profiles}
\end{figure}

\begin{figure*}
\centering
\includegraphics[width=0.85\textwidth]{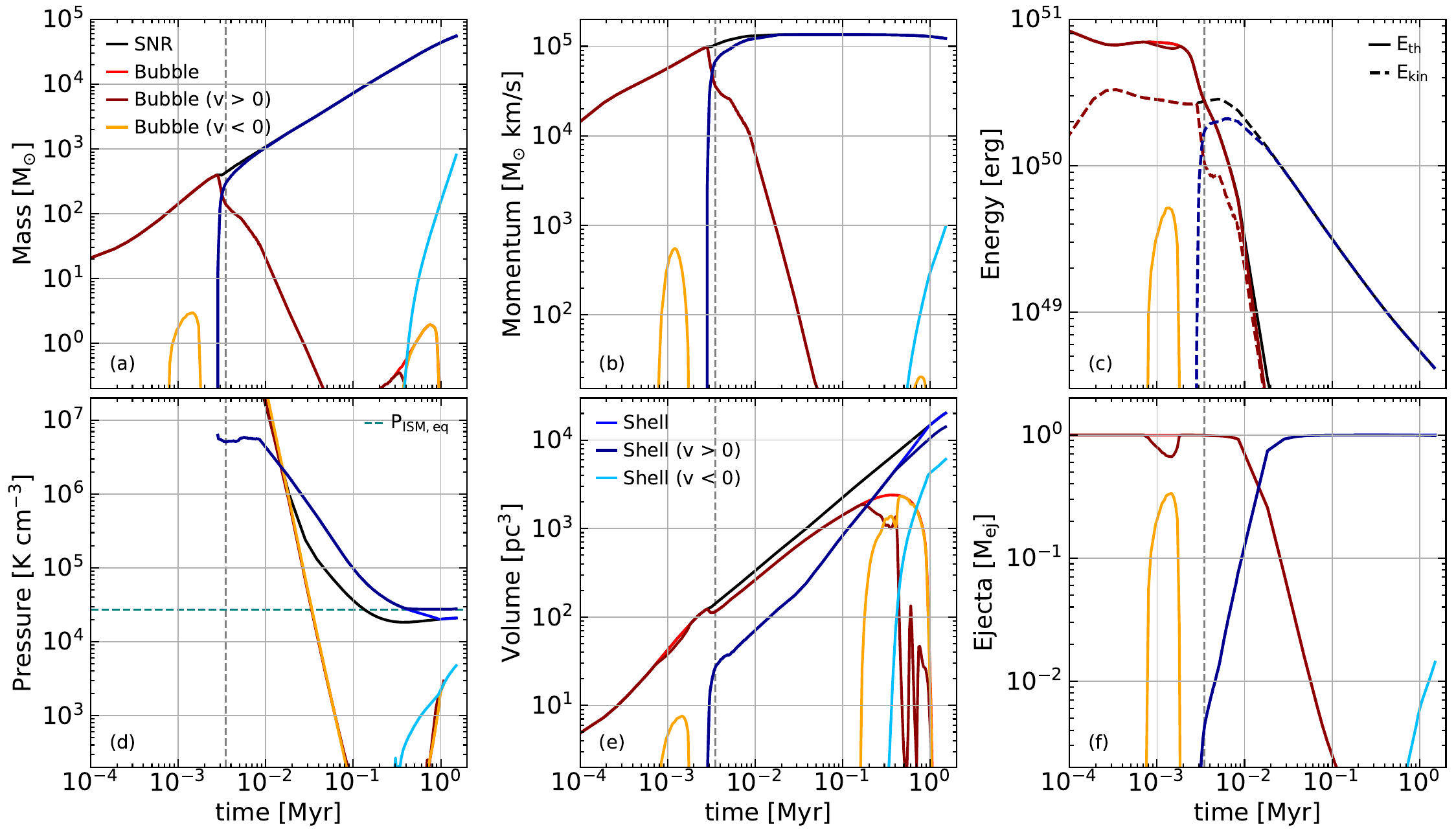}
\caption{Time evolution of various quantities for model N1\_n2\_L13: (a) mass, (b) momentum, (c) energy, (d) pressure, (e) volume, (f) ejecta. Differently colored lines correspond to different gas components as described in Figure \ref{fig:classification}. The vertical dashed line marks the theoretical estimate of the shell formation time (Equation \ref{eq:t_sf}). In panel (c) the thermal (solid) and kinetic (dashed) energy is shown. The horizontal green dashed line in panel (d) marks the equilibrium pressure of the ISM. The bubble pressure is initially very high ($\gtrsim 10^9 \, \text{K} \, \text{cm}^{-3}$) and thus outside of the axis limits.}\label{fig:global_1}
\end{figure*}

In this subsection we describe the time evolution of the SNR formed by a single SN in a stationary, uniform, high density medium in cooling equilibrium.

We first give a qualitative overview of our ultra high resolution model N1\_n2\_L14 with a maximum grid resolution of $\Delta x = 0.0625 \, \text{pc}$.
In Figure \ref{fig:slice} the density (top panels) and radial mass flux (bottom panels) is shown in slices through the $xy$-plane at four different points in time corresponding to the four different stages of SNR evolution after shell formation. 
In order to provide a quantitative reference, radial profiles of various physical quantities at the same points in time are shown in Figure \ref{fig:profiles}.

During the first stage, right after shell formation, the shell reaches a maximum compression ratio of $\chi \sim 10$. 
Its width is comparable to the resolution limit $\Delta R \sim \Delta x = 0.0625 \, \text{pc}$.
The pressure of the hot interior is about an order of magnitude higher than that of the rapidly cooling shell, and as a result bubble material is condensing onto the shell.

After 0.1 Myr, the bubble has been mostly evacuated ($n_{\text{H, Bubble}} / n_{\text{H, ISM}} < 10^{-4}$) and the shell has thickened considerably ($\Delta R \sim 3 \, \text{pc}$), leading to an overall reduction in the compression ratio ($\chi \sim 2.5$) of the shell.
The pressure in the bubble has dropped significantly to about $10\,\%$ of the ISM pressure, while the pressure of the shell is still overpressurized with respect to the ISM. 
During this stage the mass flux is concentrated within the shell, with only little outward mass flux from inside the bubble. 
The shell is subject to thin-shell overstability and nonlinear thin-shell instability \citep{1983ApJ...274..152V, 1994ApJ...428..186V, 1998ApJ...500..342B}, resulting in ripples on the shell's surface.

We note that we do \textit{not} explicitly seed perturbations that would drive these instabilities. Instead, they arise from grid scale perturbations due to the mapping of the sphere onto a Cartesian grid and numerical instabilities such as the \textit{carbuncle instability} \citep[see e.g. appendix C of][]{2008ApJS..178..137S}.
We refer the interested reader to appendix \ref{app:convergence}, where we discuss in more detail the dependence of these artifacts on the resolution and how they might affect our results.

After 1 Myr the flow just behind the shell has reversed and is now flowing inward, starting to slowly fill up and cool the bubble with material from the backside of the shell. 
Meanwhile, the pressure of the shell has dropped to a level comparable to the ISM pressure.
The shell continues to broaden ($\Delta R \sim 7 \, \text{pc}$) and the compression ratio ($\chi \sim 1.5$) continues to drop, approaching unity.
During these first three stages, most of the mass of the bubble gas is composed of ejecta material.

Finally, after 10 Myr the inward flow has reached the center and is compressed into a compact, dense, expanding cloud with a constant density of about 3 times the ISM density and a size of about 5 pc.
The mass fraction of the ejecta in the cloud is up to an order of magnitude larger than in the rest of the SNR.
The outer radius of the cloud is bounded by more inflowing material, which is slightly underdense relative to the ISM.
Meanwhile, the shell has broadened to about $\Delta R \sim 20 \, \text{pc}$ and the compression ratio is only slightly above unity.
The shell instabilities have lead to complex substructure within the shell. The shell is composed of many $\sim \text{pc}$ size \textit{blisters}, which are bounded by outflowing, overdense shells and filled with inflowing, underdense gas.

In order to describe the launching mechanism of the implosion in more detail, in Figure \ref{fig:global_1} we show the time evolution of various globally computed quantities for the model N1\_n2\_L13\_HC with a maximum grid resolution of $\Delta x = 0.125 \, \text{pc}$. 
Here we use the L13 model, because due to storage limitations it was not feasible to run an L14 model with high output cadence. The different panels show the time evolution of mass, momentum, kinetic \& thermal energy, pressure, volume and ejecta mass, calculated for the different gas components described in Figure \ref{fig:classification}.

Extensive quantities are computed by summing up the contributions from each cell belonging to the respective gas components.
The volume averaged pressure of each gas component $i$ is calculated as
\begin{equation}\label{eq:pressure}
    p_{i} = \left(\gamma-1\right) \frac{E_{\text{th}, i}}{V_{i}},
\end{equation}
where $\gamma=5/3$ is the adiabatic index, $E_{\text{th}, i}$ is the thermal energy and $V_{i}$ is the volume.

After an initial relaxation period the solution approaches the energy-conserving ST phase, during which the entire SNR is hot and $\sim 70\, \%$ of the energy is thermal, in agreement with analytical calculations \citep{1950RSPSA.201..159T, 1959sdmm.book.....S}.

Shortly before shell formation, after $t \sim 10^{-3} \, \text{Myr}$, a reverse pulse emerges, is reflected in the center and merges with the shock again.

At shell formation $t_{\text{sf}} \sim 3 \times 10^{-3} \, \text{Myr}$ the bubble mass reaches a maximum and the shell mass begins to increase. 
Similarly the momentum of the bubble gas reaches a maximum as the momentum of the shell starts to increase.
The total thermal energy begins to drop steeply, while the kinetic energy remains constant. 
The thermal energy is dominated by the bubble while the shell carries negligible amounts of thermal energy. 
On the contrary the kinetic energy of the bubble drops rapidly and is taken over by that of the shell.
The pressure of the shell after its formation is initially roughly constant and much lower than that of the bubble, which however drops rapidly.
The volume of the SNR is dominated by the hot bubble, which after shell formation initially decreases until it reaches a fixed volume filling factor of about $\sim 2/3$, with a corresponding volume filling factor of the shell of about $\sim 1/3$.
Most of the ejecta stay within the bubble and only slowly get incorporated into the shell.
This behavior is in line with the modified version of the PDS phase described in 1D by \citet{1988ApJ...334..252C} and in 3D by \citet{2015ApJ...802...99K}.

After about $10^{-2} \, \text{Myr}$ most of the SNRs mass, momentum and kinetic energy is carried by the still entirely outward moving shell.
Until this point the momentum has still been increasing, but at this point the increase stops, marking the beginning of MCS phase.
The kinetic energy starts to decrease as $t \propto t^{-0.75}$, consistent with the analytical expectation.
The pressure of the shell begins to decrease due to the effect of radiative cooling, while the pressure of the bubble keeps decreasing rapidly.
About $10\,\%$ of the ejecta are now in the shell.

After $2 \times 10^{-2} \, \text{Myr}$ the pressure of the bubble falls below that of the shell. 
At this point the volume filling factor of the bubble begins to decrease and that of the shell correspondingly has to increase, corresponding to a relative broadening of the shell.
At this point the majority of the ejecta mass is in the shell.

The backflow emerges after $t_{\text{launch}} \sim 0.3 \, \text{Myr}$ at the same time as the pressure of the shell approaches the ambient pressure.
This inward flow is quite different from the series of reflected sound waves described by \citet{1988ApJ...334..252C}.
While the sound waves are associated with the bubble and carry negligible amounts of mass, the implosion is associated with the shell and carries relatively large amounts of mass, which are growing.
The pressure of the outward moving shell levels off at the ISM pressure, while that of the inward moving component increases.
The emergence of the backflow leads to a slight decrease in total radial momentum.

After 1.5 Myr, about $1 \, \%$ of the shell mass is moving inward, carrying about $1 \, \%$ of the ejecta back to the center. 
At this point, the bubble has disappeared entirely.

\subsection{Universality of the Mechanism}\label{sec:universality}

\begin{figure*}
\centering
\includegraphics[width=0.85\textwidth]{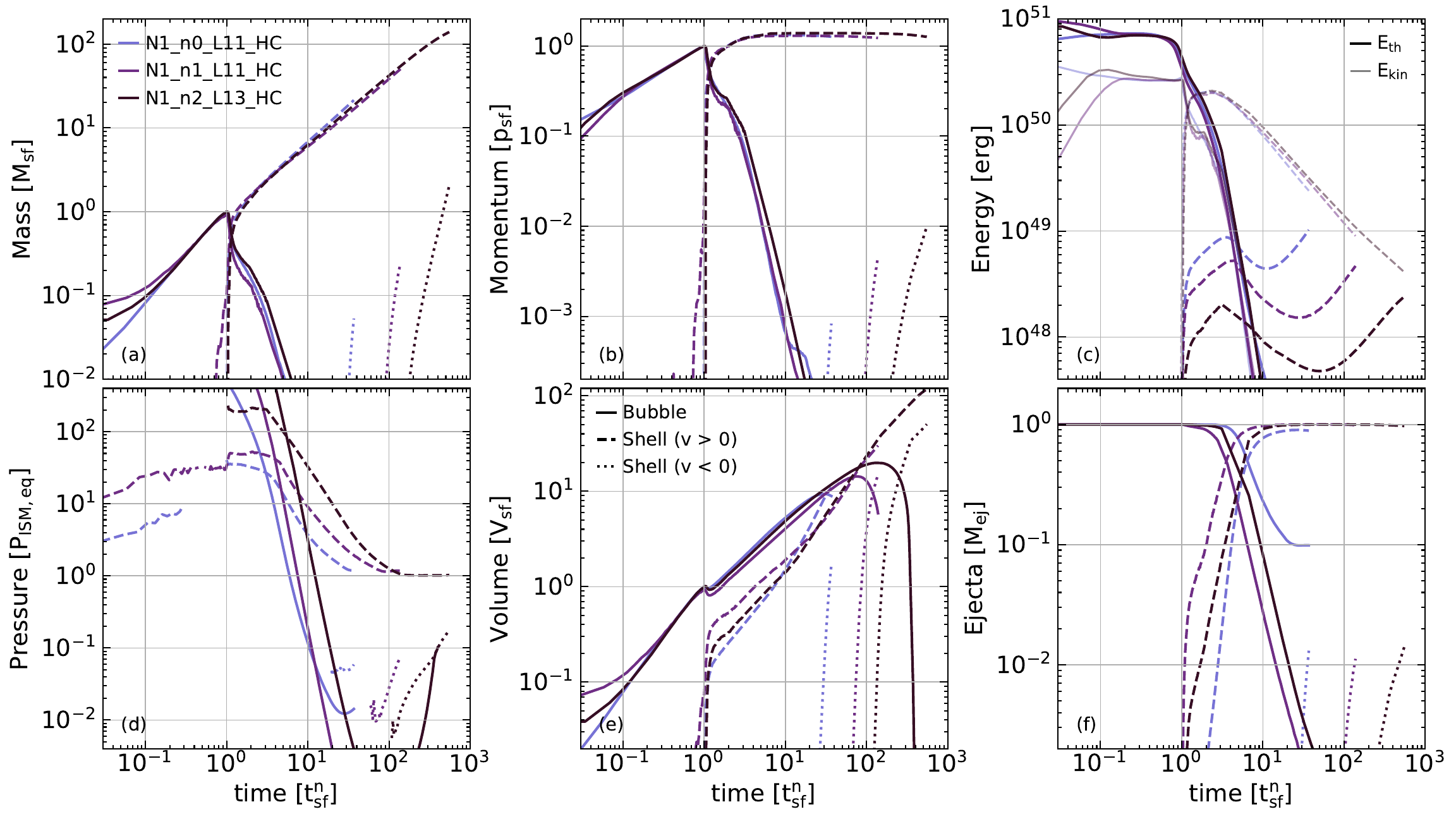}
\caption{Rescaled version of Figure \ref{fig:global_1} for various models in media with different density. Solid, dashed and dotted lines correspond to the hot bubble, outflowing, and inflowing shell, respectively. The time is normalized by $t_{\text{sf}}^{\text{n}}$ (see text). Quantities with subscript 'sf' correspond to the value of the whole SNR at $t = t_{\text{sf}}^{\text{n}}$. In panel (c) the thermal and kinetic energy are plotted with an opacity alpha of 1 and 0.5, respectively.}\label{fig:global_2}
\end{figure*}

In Figure \ref{fig:global_2} we show a rescaled version of Figure \ref{fig:global_1} for the models N1\_n0\_L11\_HC, N1\_n1\_L11\_HC and N1\_n2\_L11\_HC. Time is measured in units of the shell formation timescale $t_{\text{sf}}^{\text{n}}$. Mass, momentum and volume are normalized to their value at $t_{\text{sf}}^{\text{n}}$ and pressure is normalized to the ISM value in cooling equilibrium.

As expected from the self-similarity during the ST phase \citep{1959sdmm.book.....S}, all models exhibit a very similar time evolution before shell formation. 
In the models N1\_n0\_L11\_HC and N1\_n1\_L11\_HC there is already some small amount of cold gas during this phase. This is due to the method used to extract SNR gas, which might include a negligible number of unshocked cells.

After shell formation, the time evolution in all models is qualitatively the same as for N1\_n2\_L13\_HC, though timescales in units of the shell formation timescale may differ.
In the models N1\_n0\_L11\_HC and N1\_n1\_L11\_HC the pressure of the outward moving shell starts to decrease after the bubble pressure falls below it after 3 -- 4 $t_{\text{sf}}^{\text{n}}$, while in N1\_n2\_L13\_HC it already starts to decrease already slightly before that.

We interpret this difference as follows: The pressure can either decrease by radiative cooling or adiabatic expansion. 
When the pressure of the bubble drops below the shell pressure before radiative cooling becomes important, the shell will cool adiabatically. 
On the other hand, if the cooling timescale is shorter than the timescale to reach pressure equilibrium between the shell and the bubble, the shell will start to cool radiatively before the bubble pressure has dropped. 
In both cases the shell pressure will start to decrease, albeit with a slightly different scaling.
It is therefore no surprise that in the run with a higher density and therefore a shorter cooling timescale, the pressure starts to drop slightly earlier.

In all models, once the pressure of the shell approaches that of the ISM, a steadily growing backflow is launched.

\subsubsection{Timescales}\label{sec:timescales}

\begin{figure}
\centering
\includegraphics[width=0.45\textwidth]{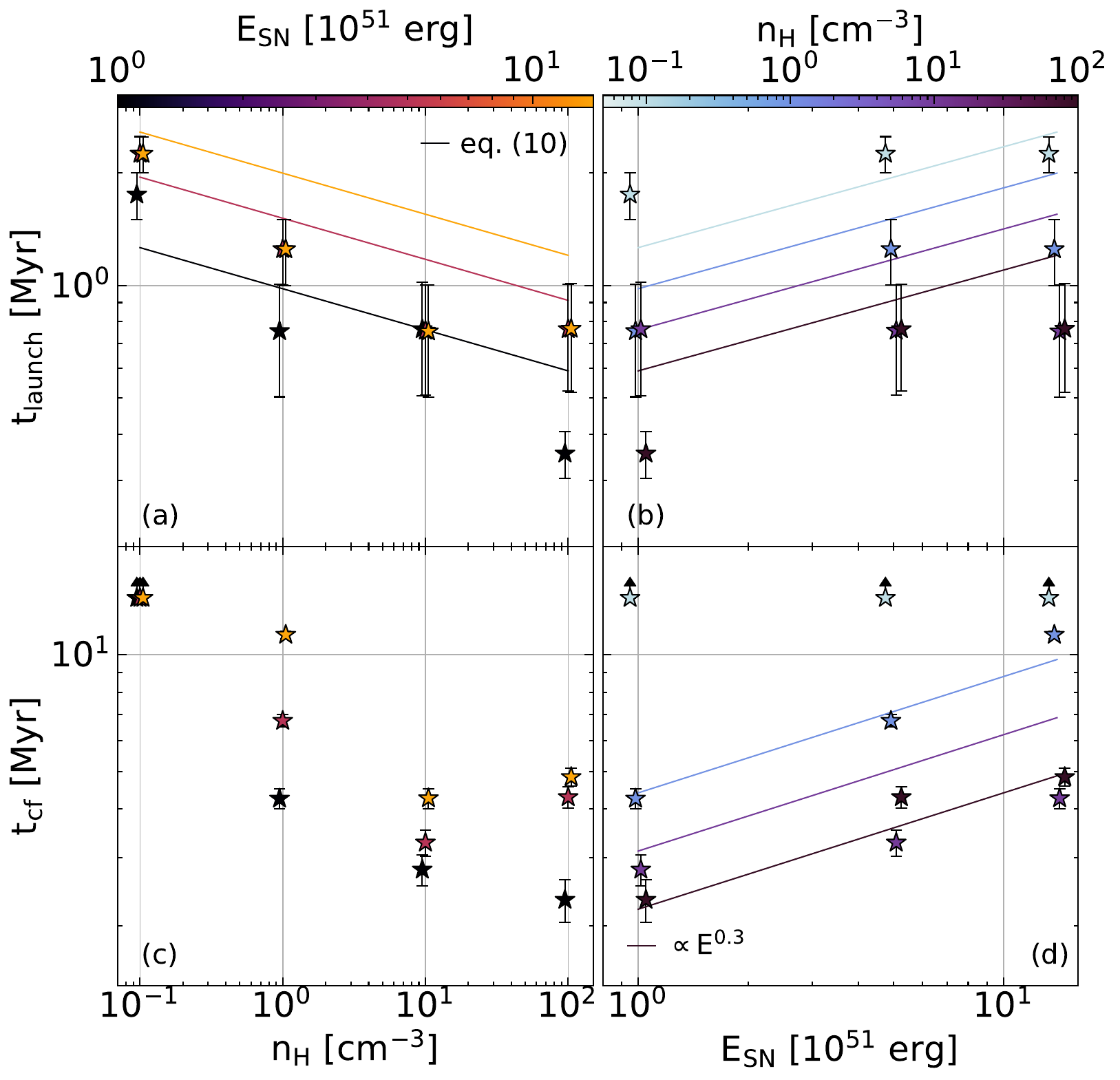}
\caption{Launching timescale (top panels, a \& b) and cloud formation timescale (bottom panels, c \& d) as a function of explosion parameters. Left panels (a \& c) show the timescales as a function of ISM density and right panels (b \& d) as a function of explosion energy. Solid lines correspond to the model described in section \ref{sec:model}. Data points are colored by the respective other explosion parameter. We added a small displacement ($\leq 5 \,\%$) to the x-values in order to reduce the overlap of the markers.}\label{fig:timescales}
\end{figure}

Having established the universal emergence of backflows, once the shell pressure approaches that of the ambient medium, we may investigate next, how the timescales for launching the backflow $t_{\text{launch}}$ and subsequently forming a central overdensity $t_{\text{cf}}$ depend on the explosion parameters.

We define the launching timescale as the earliest time, when the inflowing shell mass exceeds $0.1 \, M_{\odot}$. 
We choose this threshold, because especially in the runs, where the shell is resolved with many cells, instability of the shell itself can lead to eddies within the shell, which lead to a small amount of inflowing shell gas that is not associated to the implosion.
The cloud formation timescale is defined as the earliest time after launching, when the density in the innermost radial bin exceeds the ambient density.

Due to the limited temporal resolution of the snapshots, the events actually occur somewhere between the first snapshot when the above conditions are met and the previous snapshot. 
We therefore report the arithmetic mean of the two time points and indicate the time interval between the two snapshots with error bars.

In Figure \ref{fig:timescales} the timescales are shown as functions of the ambient density (left panels) and the explosion energy (right panels). The markers are colored by the respective other variable.

We find launching timescales between a few hundred kyr and few Myr.
There is a negative trend with density and a positive trend with energy.
The scaling and normalization are in rough agreement with Equation \ref{eq:t_launch}, with factor of $\sim 2$ differences.

At high densities differences might arise, because in the derivation of Equation \ref{eq:t_launch} we have neglected the role of radiative cooling below $T \sim 10^{4} \, \text{K}$ for the shell gas (see also discussion of Figure \ref{fig:global_2}), which would lead to a shallower scaling of $t_{\text{PDS}}$ (Eq. \ref{eq:t_PDS}) with density, which in turn manifests as a steeper scaling of $t_{\text{launch}}$.

At the low density end, the differences might arise, because here $P_{\text{shell, PDS}}$ is already quite similar to $P_{\text{ISM, eq}}$ and therefore the assumed scaling might not apply since the shock is already quite weak at $t_{\text{PDS}}$. 

The cloud formation timescale is on the order of several Myr to over 10 Myr for our simulations with the highest explosion energies and lowest densities. 
At $n_{\text{H}} = 0.1 \, \text{cm}^{-3}$ no overdense clouds have formed by the end of our simulations. 
There is no clear trend with density. 
The cloud formation timescale becomes shorter for SNRs in higher density environments, though for high explosion energies the timescale appears to level off and it even slightly increases for the highest densities.
The timescale generally increases with energy, roughly scaling like $t_{\text{cf}} \propto E_{51}^{0.3}$.

The scaling with the explosion energy appears to be the same as the scaling of the radius of the SNR at launching $R_{\text{launch}}\left(E_{51}, n_{\text{H}}\right)$, as predicted by Equation \ref{eq:R_launch}.
Equation \ref{eq:t_cf} then implies that the implosion velocity $V_{\text{in}}$ is independent of explosion energy and depends only on density. 

\newpage
\subsection{Cloud Properties}\label{sec:cloud_properties}

\begin{figure}
\centering
\includegraphics[width=0.45\textwidth]{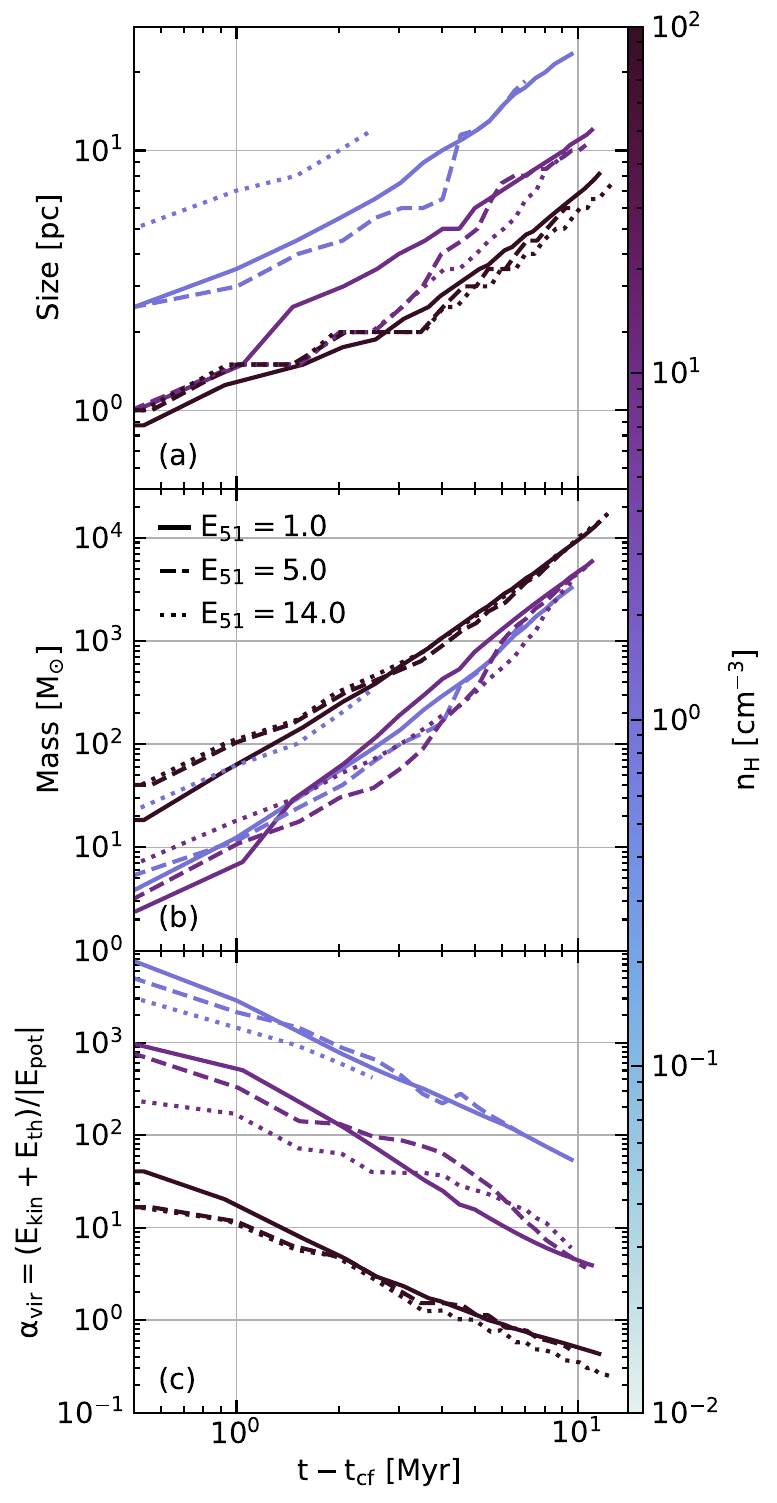}
\caption{Time evolution of the central cloud's radius (a), mass(b) and virial parameter (c) for the different models. Solid, dashed and dotted lines correspond to an explosion energy of $E_{51} = 1.0$, $E_{51} = 5.0$ and $E_{51} = 14.0$, respectively. Lines are colored by the ISM density.}\label{fig:cloud_properties}
\end{figure}

In the previous section we have shown that for a wide range of explosion parameters, the SNRs implode and form a cloud in their center. Here we summarize the properties and evolution of these clouds and discuss how they depend on the explosion parameters.

To this end, we utilize radial profiles of density, kinetic and thermal energy density to compute the size, mass and virial parameter of the clouds.
We note that our simulations are strictly without self-gravity. 
The timescale for self-gravity to have a qualitative effect on the evolution of the SNR is much longer than the simulation time.
Nonetheless, it might have an effect on the evolution of the dense cloud, and in order to estimate the importance of self-gravity for the cloud and predict whether or not it might become self-gravitating and eventually form stars it is useful to look at the virial parameter.

We define the size of the cloud as the radius of the interface between the innermost radial bin, where the density falls below the ambient density.
The cloud mass is defined as the integral of the density profile up to that radius and the virial parameter is defined as the ratio of the sum of kinetic and thermal energy and the modulus of the potential energy of the cloud. The kinetic and thermal energy are computed by integrating over the respective radial profiles and the potential energy is defined as 
\begin{equation}\label{ref:Epot}
    \left|E_{\text{pot}}\right| = \frac{3}{5} \frac{G M_{\text{cloud}}^2}{R_{\text{cloud}}}.
\end{equation}
Here we assume that the density profile within the cloud is flat and that all motion within the cloud is opposing the gravitational pull. 
Both of these assumptions approximately hold true (see e.g. Figure \ref{fig:profiles}).

In Figure \ref{fig:cloud_properties} we present the time evolution of the cloud's size, mass and virial parameter for the different models.

In panel a) the cloud radius is shown as a function of time.
The cloud grows to a size of several to few tens of parsecs within 10 Myr.
In lower density environments clouds grow larger, with little dependence on the explosion energy.

Panel b) shows the cloud's mass as a function of time.
At the end of the simulation, the mass of the central cloud has reached a value of several thousand to about $2 \times 10^4$ solar masses, with only a small dependence on the density of the ambient medium.
Clouds in high density environments become somewhat more massive.
There is only little dependence on the explosion energy.

Panel c) shows the cloud's virial parameter as a function of time.
The initial virial parameter scales with the ambient density roughly as
\begin{equation}\label{eq:virial}
    \alpha_{\text{vir,0}} \sim 10^4 \, n_{0}^{-1},
\end{equation}
where $n_{0} = n_{\text{H}} / \text{cm}^{-3}$, and decreases steadily approaching and dropping below $\alpha_{vir} \sim 1$ for the high density runs, suggesting that indeed the clouds would become self-gravitating.
There is little variation due to the explosion energy.

The masses and radii of the clouds formed by SN implosion are comparable to the values used in the initial conditions for the clouds in the STARFORGE \citep{2022MNRAS.512..216G, 2023arXiv230700052G, 2023arXiv230911415F} simulation suite, which studies the star formation from the collapse of a single giant molecular cloud. This suggests, that the thus formed clouds might indeed trigger another generation of star formation.

\section{Discussion}\label{sec:discussion}

In the previous section we have described a new mechanism by which a cloud can form inside a radiative SNR due to its implosion as the shell pressure approaches that of the ISM.
In the following, we will discuss some of the limitations of our model, the role of some of our model ingredients and the implications of our findings in the context of galaxy evolution.

\subsection{Limitations}\label{sec:limitations}

In this work, we have shown the existence of SN implosions and subsequent cloud formation in the center of the SNR, using a suite of hydrodynamic simulations of SN explosions in a uniform and stationary medium. 
Of course, a uniform and stationary ISM is a great simplification of the complexity of a realistic ISM. 

In a more realistic model for the ISM, like the kind of turbulent, stratified box used in state-of-the-art ISM simulations \citep[see e.g.][]{2015MNRAS.454..238W, 2017ApJ...846..133K}, the existence of backflows as described in our work is a priori not guaranteed. 
Continuous or sufficiently frequent energy injection can keep the SNR overpressurized and prevent an implosion \citep{2017ApJ...834...25K}. 
Indeed, stellar populations are expected to explode SNe in regular intervals before they run out of fuel \citep[see e.g.][]{1999ApJS..123....3L}. 
However, in the cases of runaway stars or populations hosting a sufficiently small number of massive stars this limitation does not apply and even in the case of stellar populations that remain active for a long time, eventually the bubbles are going to evacuate and cool off enough to make an implosion feasible.

Besides the importance of continuous driving, the role of the ambient medium cannot be ignored. In a more realistic description of the ISM, the ambient medium is highly structured due to the combined effect of turbulence, shear and stratification. 
SNe exploding in such an environment will follow the geometry of the ISM \citep{2023MNRAS.523.1421M}, as the shock wave can only slowly penetrate into dense structures, but will quickly fill out the volume filling low density medium, leading to highly amorphous SNR shapes \citep{2015ApJ...802...99K, 2021ApJ...914...90L}.
In such a configuration, the SNR will reach pressure equilibrium at different times in different directions, leading to a displacement and deformation of the clouds formed in this way. 
If the momentum carried by the backflowing gas from different directions is not equal and opposite, the cloud would further end up with a net momentum leading to a drift.
Similar asymmetries can follow from the interaction of the SNR with neighboring shocks, e.g. due to neighboring superbubbles \citep{2000A&A...361..303B} or small-scale turbulence, which can locally contribute to the pressure opposing the shock expansion. In Appendix \ref{app:triggering} we show, that asymmetries in the ambient pressure can indeed trigger an implosion locally.

Besides the limitations due to the environment, more complete physics might also qualitatively modify our conclusions.

As mentioned in section \ref{sec:cloud_properties} the virial parameter of the clouds drops below unity a few Myr after their formation. 
While it seems plausible that self-gravity is too weak to have an important effect on the implosion mechanism and the crossing of the inflowing gas, once the cloud has formed, it might collapse and fragment due to its self-gravity.

\citet{2015ApJ...802...99K} have demonstrated that magnetic fields play only a subdominant role in SNR evolution. 
On the contrary, \citet{2019MNRAS.483.3647G} show that magnetic fields suppress the growth of instabilities at the bubble-shell interface, which reduces mixing and delays radiative cooling in the case of multiple consecutive SNe. 
However, it is important to note that \citet{2015ApJ...802...99K} used a mesh code, while \citet{2019MNRAS.483.3647G} used a Lagrangian method. 
While both methods achieve comparable resolution in the shell, the mass resolution in the bubble is several orders of magnitude higher in mesh codes, which are thus likely less affected by spurious cooling of bubble gas.
Nonetheless, contributions to the pressure from a magnetic field and cosmic rays can potentially alter the timescales for implosion and cloud formation or even prevent these processes, if they can maintain a high enough bubble pressure.
Indeed, in the case of the hydromagnetic RTI \citet{2000A&A...361..303B} have shown, that the magnetic field has a stabilizing effect, that implies a characteristic length scale at which the instability can act. In this case the backflowing gas fragments into blobs of size similar to the fastest growing wavelength.

\citet{2014MNRAS.443.3463S} conducted 1D spherically symmetric simulations of SNe. 
They note that most of the heat losses occur in the unresolved layer between the bubble and the shell. 
The physical width of this layer is much too small to resolve, even with their 1D method, but they show that nonetheless the cooling losses are converged, even for moderate resolution.

Similarly, as discussed by \citet{2016MNRAS.456..710F}, while the length scale associated with thermal conduction is much too small to be resolvable with current techniques, its effect is negligible.
However, recent results by \citet{2019MNRAS.490.1961E} in the context of continuously driven superbubbles suggest that heat conduction does in fact play an important role for the transport of energy and mass across the bubble-shell interface. 
However, it is worth nothing that \citet{2019MNRAS.490.1961E} artificially enhanced the conduction rate to model turbulent mixing due to 3D instabilities, which makes a direct comparison difficult. 
\citet{2021ApJ...914...90L} simulate the expansion of a continuously driven wind bubble in a turbulent medium confirming that turbulence indeed enhances the mixing across the bubble-shell interface leading to catastrophic cooling losses in ideal hydrodynamics.
Yet, the importance of these effects remains unclear in a picture where magnetic fields suppress the growth of the instabilities responsible for the mixing.
Further studies of individual SNe with resolved heat conduction, magnetic fields and turbulence are required in order to settle this ongoing debate. 

\subsection{Role of the cooling model}\label{sec:cooling}

In section \ref{sec:model} we have presented a model for the launching of the backflow. There are two ingredients of our model that are sensitive to assumed cooling physics.
First, we have assumed that the temperature of the shell right after shell formation remains stable at $10^4 \, \text{K}$, as the time it takes to cool beyond this temperature is much longer than the dynamical timescale.
Second, we assume that the ISM is in cooling equilibrium, when we equate the shell pressure with the ISM pressure.
Both of these assumptions invoke the assumed cooling physics, which may affect the resulting timescales and cloud properties. 

\citet{2023ApJS..264...10K} compare their detailed radiative transfer model to a range of commonly used cooling functions (see their Figure 17). 
They find that in these functions the equilibrium pressure at a given density may differ by up to three orders of magnitude between the models. 
In the \textsc{Ramses} cooling model utilized in this work the equilibrium pressure is indeed relatively high.
As a consequence the SNR is expected to implode 100 times earlier with our cooling model, than e.g. with the model by \citet{2020MNRAS.497.4857P}, which has an exceptionally low equilibrium pressure.
Further complications like thermal instability and non-equilibrium effects \citep{2022ApJ...938...23K}, might also qualitatively alter our conclusions.

For a more detailed discussion, we refer the interested reader to appendix \ref{app:cooling}, where we compare the results of simulations with different cooling models.
The comparison indicates, that indeed a lower ambient pressure will delay the implosion and that the details of the cooling physics, may have a slight effect on the details of cloud formation, even at a comparable ambient pressure.

\subsection{SN Implosion in the Literature}\label{sec:literature}

Despite the fact that the evolution of SNRs in a uniform medium has been studied in great detail for more than 40 years, to our knowledge there has been no mention of SN implosion.
Here we discuss the various reasons for why this process might have remained unnoticed for so long.

Many authors \citep[e.g.][]{1974ApJ...188..501C, 1974ApJ...190...59S, 1998ApJ...500...95T, 2015ApJ...802...99K} only focus on the transition to the radiative phase and would therefore not advance their simulations far enough to reach the implosion stage.

\citet{1988ApJ...334..252C} set the pressure of the ambient medium to an artificially low value in order to maintain a strong shock, which in turn delays the implosion.

\citet{2016MNRAS.456..710F} use one-dimensional hydrodynamic simulations to study the energy input from SNe in a uniform and stationary media with a range of densities and an initial temperature of 1000 K. They advance their models until the shock velocity approaches the sound speed of the ISM and thus in principle should have been able to see an implosion. However, the internal structure of the SNRs was out of their scope and thus they did not report any backflow.

\citet{2020GApFD.114...77G} utilize three-dimensional hydrodynamic simulations of radiative SNRs to validate the PENCIL code. 
Even though they advance their models until the shock becomes sonic, as they mostly focus on benchmarking their code with previous work, they do not report any backflow.

\citet{2000A&A...361..303B} use linear perturbation theory to describe a type of hydromagnetic RTI that, in the limit of vanishing magnetic field, is very similar in nature to the SN implosion described here.
They show, that the interaction of two SNRs can lead to an inward flow of clouds originating from the interaction region.
Our results, which appear to correspond to the same kind of instability, seem to confirm the linear prediction of \citet{2000A&A...361..303B}.

To our knowledge there are no instances of SN implosions reported in observations.
This might however simply attest to the fact, that radiative SNRs, in particular those close to merging with the ISM, are very dim and thus are often difficult to observe \citep{2019JApA...40...36G, 2020ApJ...905...35K, 2023arXiv230803484Z}. 
Moreover, the fact that the morphology of imploding SNRs differs qualitatively from traditional SNRs might have lead to a misclassification of imploding SNRs as something other than a SNR.

\newpage
\subsection{Implications for Galaxy Evolution} \label{sec:implications}

We have shown that under quite general conditions, old SNRs will implode and form a compact, massive, and potentially self-gravitating cloud in its center.
Such a cloud could collapse and fragment under its own self-gravity to form stars, suggesting a novel mode of positive feedback.

Furthermore, given that such clouds would be highly enriched with the SN ejecta, and therefore with short-lived radionuclides (SLRs) like $^{26}\text{Al}$, this provides an attractive pathway to the formation of planetary systems, where the heating due to SLRs plays an important role \citep{1955PNAS...41..127U}. Indeed, it has been concluded by \citet{2021NatAs...5.1009F} that the enrichment with SLRs would have to occur prior to core formation, a condition that at face value is readily fulfilled by our proposed mechanism. However, further studies are necessary to investigate to what extent mixing due to small-scale turbulence might further dilute the imploding gas.

Besides the importance for star and planet formation we make a clear prediction for the lifetime of hot cavities in the ISM, which are filled shortly after the implosion is launched.
This number is an important parameter in models for the multiphase structure of the ISM \citep{1977ApJ...218..148M, 2003ApJ...587..278W, 2011piim.book.....D}, which are used to estimate a wide variety of ISM properties.

\section{Concluding Remarks}\label{sec:summary}

We have performed 3D hydrodynamic simulations of SNRs in a uniform, stationary medium with non-negligible thermal pressure in order to study their evolution after shell formation.
Our simulations reveal that radiative SNRs implode after the shell reaches pressure equilibrium with the ISM. 
The implosion leads to the formation of a compact, massive cloud that might soon become self-gravitating and that is highly enriched with SN ejecta. 
As we discuss, this novel mechanism of cloud formation provides attractive initial conditions for star and planet formation and might have some important implications for the theory of the ISM.

While the idealized setup is useful for understanding the underlying physical mechanism, understanding the role of SN implosion and subsequent cloud formation in a more realistic setup deserves further investigation.

We conclude that the dispersal and merging of SNRs with the ISM offers a wealth of hidden complexities, which deserve further study as they can help understand the physics of the ISM.

\begin{acknowledgments}
We are grateful to Guang-Xing Li for fruitful discussions.
We thank the anonymous referee for their insightful comments and suggestions that helped to improve the quality of this work.
Computations were performed on the HPC system Cobra at the Max Planck Computing and Data Facility.
This research was funded by the Deutsche Forschungsgemeinschaft (DFG, German Research Foundation) under Germany's Excellence Strategy – EXC 2094 – 390783311.
\end{acknowledgments}

\vspace{5mm}


\software{\textsc{Julia} v1.6.5 \citep{2014arXiv1411.1607B},
          \textsc{Matplotlib} v3.5.1 \citep{thomas_a_caswell_2021_5773480},
          \textsc{Mera} v1.4.0 \citep{Mera},
          \textsc{Paraview} v5.11.1 \citep{Paraview},
          \textsc{Ramses} v19.10 \citep{2002A&A...385..337T}
          }



\appendix

\begin{figure*}
\centering
\includegraphics[width=0.85\textwidth]{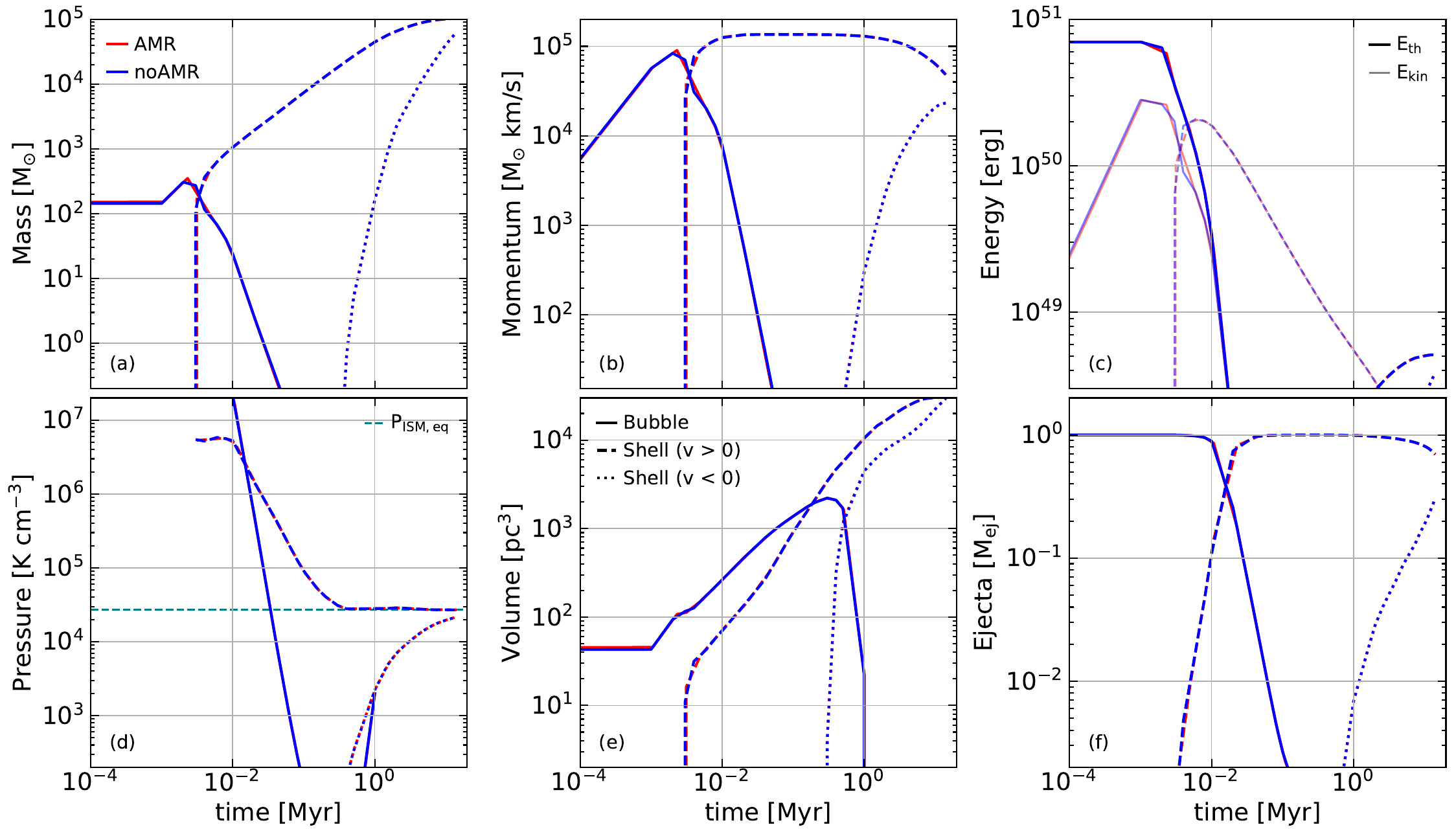}
\caption{Similar to Figure \ref{fig:global_1} for model N1\_n2\_L13 with and without AMR. Solid, dashed and dotted lines correspond to the hot bubble, outflowing and inflowing shell, respectively. In panel (c) the thermal and kinetic energy are plotted with an opacity alpha of 1 and 0.5, respectively.}\label{fig:app_amr}
\end{figure*}

\section{Adaptive Mesh Refinement} \label{app:amr}

In our fiducial simulation suite presented in the main body of the paper we have used AMR to reduce the numerical cost of the simulations. 
In our prescription we only refine cells, which have been sufficiently enriched by SN ejecta.
We thus create a central refinement region that should expand at roughly the same rate as the shock.
However, as the ejecta are physically confined behind the contact discontinuity, the refinement region might lag behind the shock and thus one might expect the shock to be slightly less refined than the rarefied gas behind it.

In SNRs the contact discontinuity and the shock are essentially at the same location and thus we expect this lag to be negligible, especially considering the presence of numerical diffusion, which acts to smear out the contact discontinuity.

In order to test whether the use of AMR can qualitatively modify our results in Figure \ref{fig:app_amr} we compare the results of the models N1\_n2\_L13 and N1\_n2\_L13\_noAMR.
The two models are almost identical, but while in N1\_n2\_L13 we use our AMR prescription in N1\_n2\_L13\_noAMR we use a static mesh with a grid spacing equal to the finest spacing in N1\_n2\_L13.

Figure \ref{fig:app_amr} shows that the two models are essentially identical. Negligibly small differences arise due to small differences in the time when snapshots are written, since they are only written when the entire grid is synchronized.

We thus conclude that the adopted AMR technique does not affect our results in any meaningful way.

\section{Convergence} \label{app:convergence}

\begin{figure*}
\centering
\includegraphics[width=0.85\textwidth]{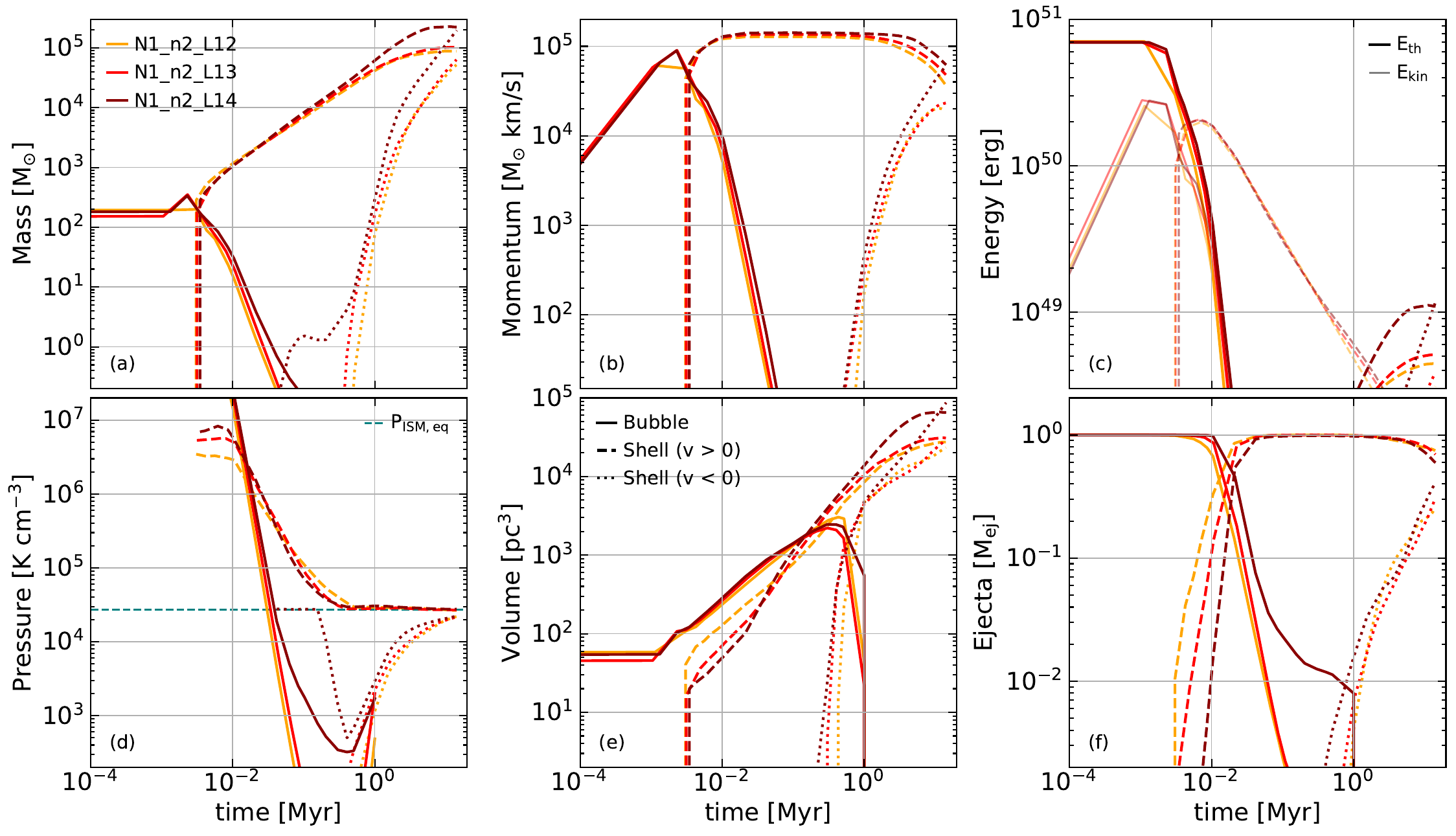}
\caption{Similar to Figure \ref{fig:global_1} for models with different resolution. Solid, dashed and dotted lines correspond to the hot bubble, outflowing and inflowing shell, respectively. In panel (c) the thermal and kinetic energy are plotted with an opacity alpha of 1 and 0.5, respectively.}\label{fig:app_res}
\end{figure*}

In order to test, whether our results are converged, we ran the N1\_n2 model at three different resolutions, denoted by N1\_n2\_L12, N1\_n2\_L13 and N1\_n2\_L14.
The results are shown in Figure \ref{fig:app_res}. Shown are the mass, momentum, energy, pressure, volume and ejecta content of the different gas components as a function of time.

Before shell formation, there are only minor differences between the models, which seem to stem from slightly different snapshot times and the analysis procedure.
There is a slight dip in the mass, momentum and energy of the hot phase, in the lowest resolution run, which likely might have arisen from the SNR selection criterion (Figure \ref{fig:classification}).

The shell formation timescale, as well as the properties of the SNR at shell formation appear converged in line with the convergence criteria by \citet{2015ApJ...802...99K}.

Slight differences appear after shell formation. 
The pressure of the cold phase, during the PDS increases with resolution. 
Since the temperature of the shell during this phase is roughly fixed to about $10^4 \, \text{K}$, the pressure is determined by the density of the shell, which depends on the width of the shell. 
In all runs, the shell during the PDS is resolved with only few resolution elements and thus indeed is not converged, as already noted by \citet{2014MNRAS.443.3463S} who estimate that the width of the shell should be $\sim 0.001 \, R_{\text{sf}} \sim 10^{-3} - 10^{-2}\, \text{pc}$.
Despite the different PDS shell pressure, the PDS ends at a similar time in all runs and the shell pressure approaches a very similar evolution during the MCS and following phases. 

In the lower resolution runs more of the ejecta tend to be incorporated in the shell right after shell formation. However, these differences are negligible by $t = 0.1 \, \text{Myr}$.

During the MCS, prior to implosion, there is already a small non-growing fraction of backflowing cold gas in the highest resolution run. 
This component is in pressure equilibrium with the ISM suggesting that it arises from small blisters in the shell's outer edge, which form, when the shell fragments due to thin shell shell instabilities which are unresolved in the lower resolution runs.

The implosion is launched at roughly the same time after $\sim 300 - 500 \, \text{Myr}$ and forms a central cloud ($V_{\text{hot}} \rightarrow 0$) after 1 Myr in all runs, indicating that these timescales are converged.

The mass, momentum, thermal energy and volume of the cold components diverge at late times.
Despite this non-convergence all SNRs exhibit the same qualitative features, and thus it does not affect our conclusions in any meaningful way.

\section{Cooling Models} \label{app:cooling}

\begin{figure}
\centering
\includegraphics[width=0.4\textwidth]{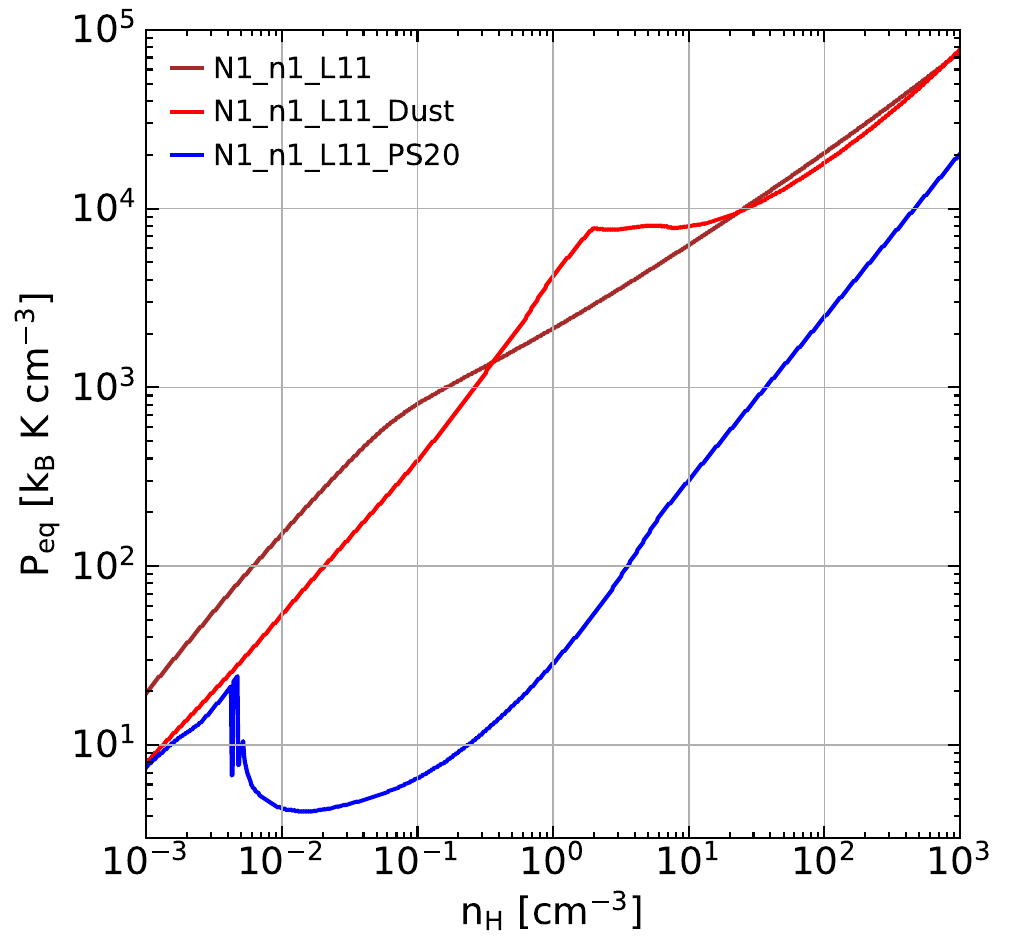}
\caption{Equilibrium pressure as a function of density for different cooling models. The brown curve corresponds to the default \textsc{Ramses} cooling, and the red and blue curves correspond to different models taken from \citet{2020MNRAS.497.4857P}.}\label{fig:app_eqlbr}
\end{figure}

\begin{figure*}
\centering
\includegraphics[width=0.85\textwidth]{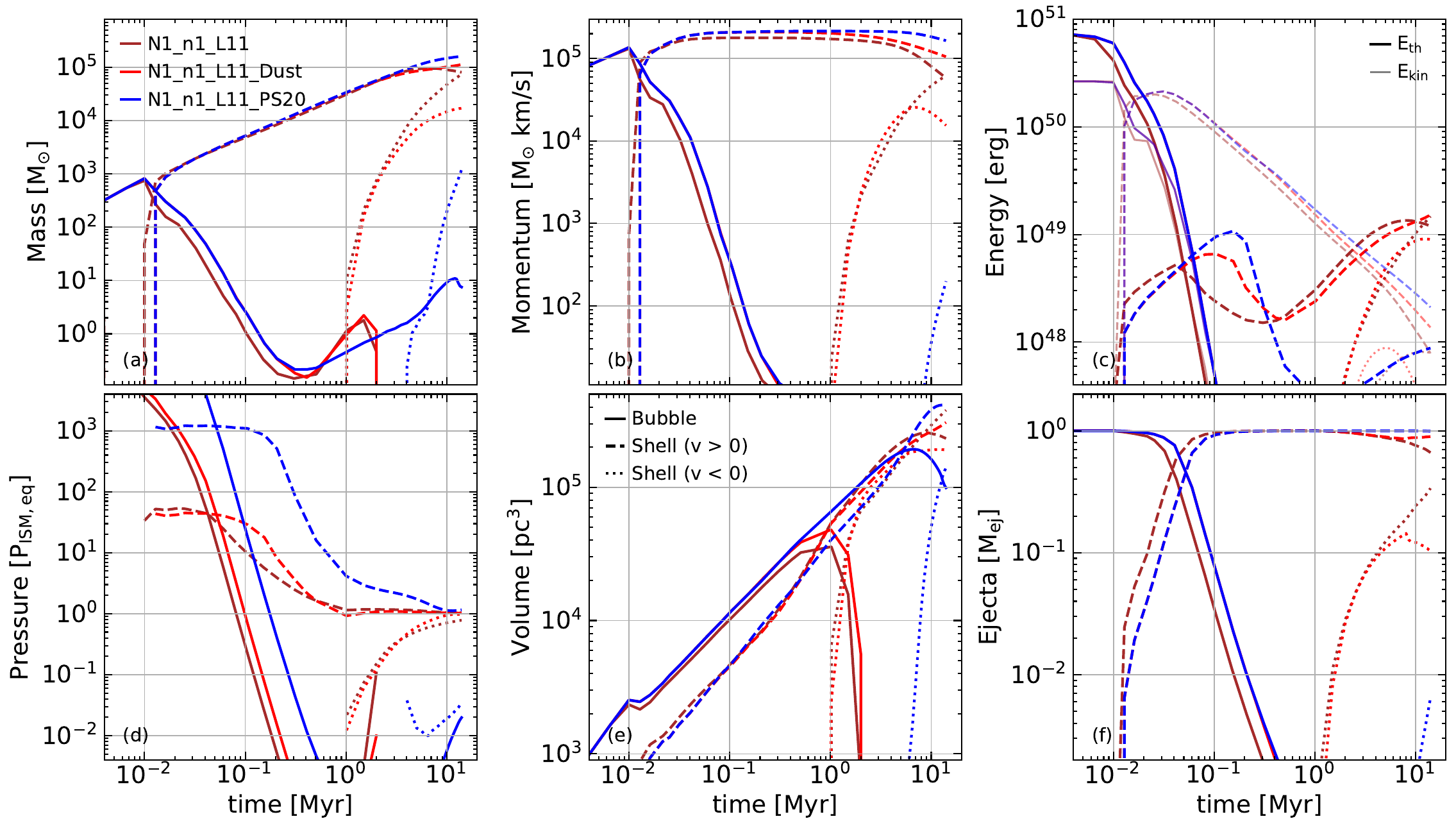}
\caption{Rescaled version of Figure \ref{fig:global_1} for models with different cooling functions. Solid, dashed and dotted lines correspond to the hot bubble, outflowing and inflowing shell, respectively. In panel (c) the thermal and kinetic energy are plotted with an opacity alpha of 1 and 0.5, respectively.}\label{fig:app_cool}
\end{figure*}

In order to investigate the role of the cooling function, we have rerun the model N1\_n1\_L11 with two different cooling models taken from \citet{2020MNRAS.497.4857P} integrated using an exact integration scheme \citep{2009ApJS..181..391T, 2017MNRAS.470.1017Z}. 
For details of the implementation we refer the reader to Behrendt et al. (in prep.). 
The cooling functions in the models N1\_n1\_L11\_Dust (\textsc{Dust}) and N1\_n1\_L11\_PS20 (\textsc{PS20}) correspond to their models UVB\_dust1\_CR0\_G0\_shield0 and UVB\_dust1\_CR1\_G1\_shield1, respectively. 
The models' cooling-equilibrium curves in the $P-n_{\text{H}}$-plane are shown in Figure \ref{fig:app_eqlbr}.

\textsc{Dust} differs from the \textsc{Ramses} cooling model in that the equilibrium curve has a pronounced kink at around $n_{\text{H}} \sim 1 \, \text{cm}^{-3}$, while \textsc{PS20} has a steep drop in pressure at around $n_{\text{H}} \sim 0.01 \, \text{cm}^{-3}$ above which the pressure is 2-3 orders of magnitude below the pressure in the \textsc{Ramses} model.

In Figure \ref{fig:app_cool} we show a comparison of the results for the different models. 
As expected the SNR evolution before shell formation is not affected by the cooling. 
The length of the PDS phase and the momentum boost during this phase are slightly increased 
for the \textsc{Dust} and \textsc{PS20} models.

Despite these differences, the timescales for implosion and cloud formation hardly differ between the \textsc{Ramses} and \textsc{Dust} models. 
On the contrary, for the \textsc{PS20} model, since the equilibrium ISM pressure is several orders of magnitude lower, implosion is delayed by $\gtrsim 3 \, \text{Myr}$ and does not lead to cloud formation within the simulated time.

These results indicate that it is indeed the ISM pressure, which controls the implosion, since all other quantities that might have an effect do not differ very much between the runs.

\section{Locally Triggered Implosions} \label{app:triggering}

\begin{figure*}
\centering
\includegraphics[width=0.85\textwidth]{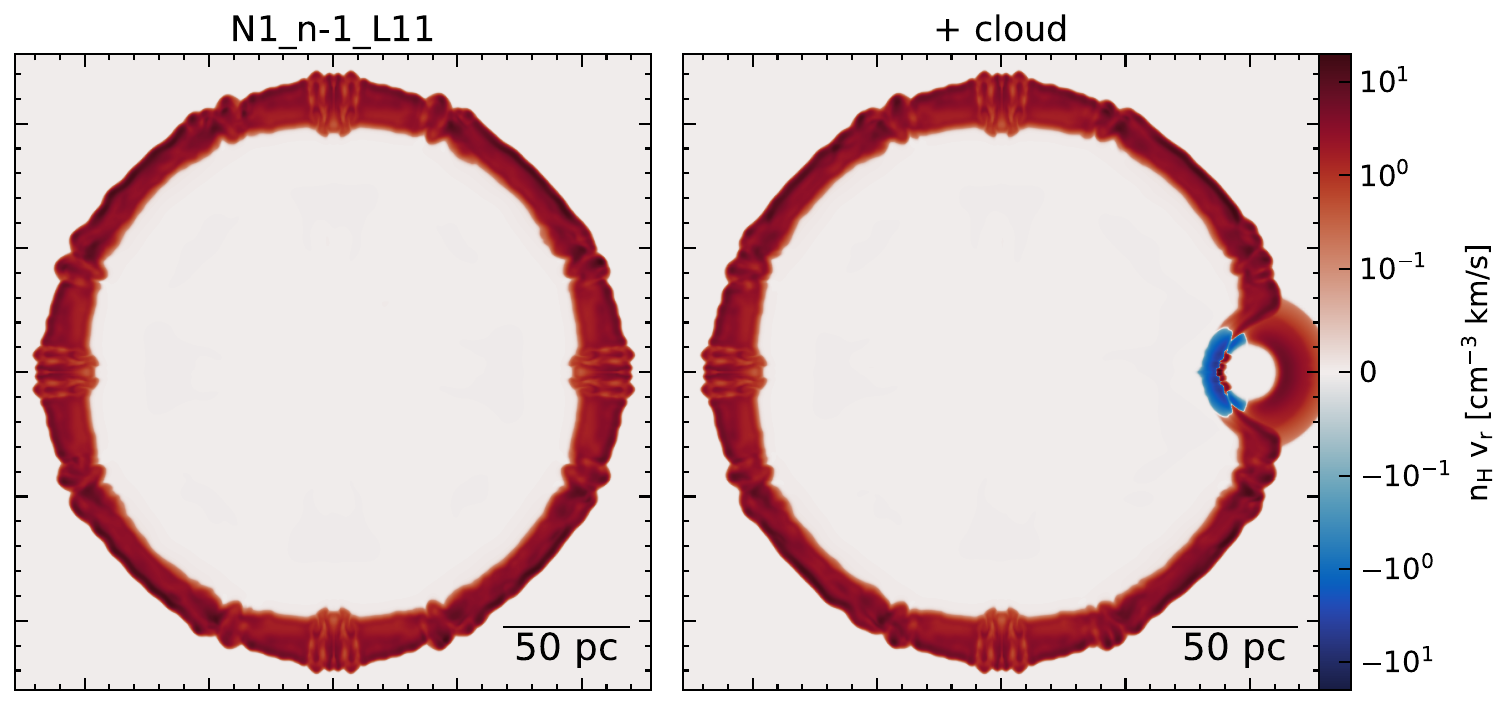}
\caption{Slices through the XY-plane of the radial mass flux 1.5 Myr after the explosion for models with and without a cold, dense cloud placed near the explosion center.}\label{fig:app_triggered_implosion}
\end{figure*}

In order to investigate, whether or not the implosion can be triggered by local pressure enhancements, we have rerun the model N1\_n-1\_L11 for 1.5 Myr with a cold ($T \sim 800 \, \text{K}$), dense ($n_{\text{H}} \sim 10 \, \text{cm}^{-3}$) cloud of Radius $R_{cl} = 15 \, \text{pc}$ centered at a distance of $d_{cl} = 100 \, \text{pc}$ from the explosion center. The pressure in the cloud is about an order of magnitude higher than that of the ambient medium.

As shown in Figure \ref{fig:timescales}, N1\_n-1\_L11 does not implode until $t = 1.75 \pm 0.25 \, \text{Myr}$ and thus there should be no \textit{global} implosion within the simulated time. However, if the implosion is indeed a local effect, the increased pressure within the cloud, should trigger a \textit{local} implosion shortly after impact.

In Figure \ref{fig:app_triggered_implosion} we show slices of the radial mass flux, 1.5 Myr years after the explosion for the runs with and without a cloud. 
In the run without a cloud there is no implosion, while in the run with the cloud, there is a significant backflowing component behind the shock coming from the direction of the cloud.
This confirms that the implosion can indeed be triggered \textit{locally}.

We note, that the cloud is slowly expanding due to the pressure gradient relative to the ambient medium, leading to a radially inflowing component downstream of the shock. This flow is unrelated to the implosion, which is necessarily upstream.


\bibliography{bibliography}{}
\bibliographystyle{aasjournal}



\end{document}